\documentclass[journal,12pt,onecolumn,draftclsnofoot,]{IEEEtran}
\pdfoutput=1
\usepackage{amsfonts,amsmath,amssymb}
\usepackage{algorithm}
\usepackage{algorithmic}
\usepackage{graphicx}
\usepackage{subfigure}
\usepackage{float}
\usepackage{diagbox}
\usepackage{makecell}
\usepackage{multirow}
\usepackage{threeparttable}
\usepackage{color}
\usepackage{xcolor}
\usepackage{xpatch}
\usepackage{hyperref}
\usepackage[hyphenbreaks]{breakurl}
\begin{document}

\title{Deep CSI Compression for Dual-Polarized Massive MIMO Channels with Disentangled Representation Learning}

\author{Suhang~Fan,
        Wei~Xu,~\IEEEmembership{Senior Member,~IEEE},
        Renjie~Xie,~\IEEEmembership{Member,~IEEE},
        Shi~Jin,~\IEEEmembership{Fellow,~IEEE},
        Derrick~Wing~Kwan~Ng,~\IEEEmembership{Fellow,~IEEE},
        and~Naofal~Al-Dhahir,~\IEEEmembership{Fellow,~IEEE}

\thanks{S. Fan and W. Xu are with the National Mobile Communications Research Laboratory, Southeast University, Nanjing 210096, China, and also with the Purple Mountain Laboratories, Nanjing 211111, China (e-mail: shfan@seu.edu.cn; wxu@seu.edu.cn).}
\thanks{R. Xie is with the School of Internet of Things, Nanjing University of Posts and Telecommunications, Nanjing 210003, China (e-mail: renjie\_xie@njupt.edu.cn).}
\thanks{Shi Jin is with the National Mobile Communications Research Laboratory, Southeast University, Nanjing 210096, China (e-mail: jinshi@seu.edu.cn).}
\thanks{D. W. K. Ng is with the School of Electrical Engineering and Telecommunications, University of New South Wales, Sydney, NSW 2052, Australia (e-mail: w.k.ng@unsw.edu.au).}
\thanks{N. Al-Dhahir is with the Department of Electrical and Computer Engineering, University of Texas at Dallas, Richardson, TX 75080 USA (e-mail: aldhahir@utdallas.edu).}
\thanks{The source code of this paper is available on GitHub: \url{https://github.com/SEUshf/DiReNet}.}
}

\maketitle
\vspace{-1.5cm}
\begin{abstract}
Channel state information (CSI) feedback is critical for achieving the promised advantages of enhancing spectral and energy efficiencies in massive multiple-input multiple-output (MIMO) wireless communication systems. Deep learning (DL)-based methods have been proven effective in reducing the required signaling overhead for CSI feedback. In practical dual-polarized MIMO scenarios, channels in the vertical and horizontal polarization directions tend to exhibit high polarization correlation. To fully exploit the inherent propagation similarity within dual-polarized channels, we propose a disentangled representation neural network (NN) for CSI feedback, referred to as DiReNet. The proposed DiReNet disentangles dual-polarized CSI into three components: polarization-shared information, vertical polarization-specific information, and horizontal polarization-specific information. This disentanglement of dual-polarized CSI enables the minimization of information redundancy caused by the polarization correlation and improves the performance of CSI compression and recovery. Additionally, flexible quantization and network extension schemes are designed. Consequently, our method provides a pragmatic solution for CSI feedback to harness the physical MIMO polarization as a priori information. Our experimental results show that the performance of our proposed DiReNet surpasses that of existing DL-based networks, while also effectively reducing the number of network parameters by nearly one third.
\end{abstract}


\IEEEpeerreviewmaketitle
\section{Introduction}

\IEEEPARstart{M}{assive} multiple-input multiple-output (MIMO) has emerged as a key technology for unleashing the potential of the fifth-generation (5G) wireless communication systems \cite{MIMO_5g}. By deploying numerous multiple antennas at both the transmitter and receiver, massive MIMO enables achieving multiplexing and diversity gains for effectively utilizing spatial resources, enhancing channel capacity, and reducing multiuser interference, under limited spectral resources and transmit power \cite{MIMO_advantage1}, \cite{RI6G}. However, to fully realize these promised advantages, it is imperative for the base station (BS) to accurately acquire the channel state information (CSI) for both the uplink and downlink channels \cite{MIMO_CSI}. For practical systems adopting the frequency division duplex (FDD) protocol, the uplink and downlink operate at different frequency bands such that channel reciprocity is absent \cite{FDD_reci}. As such, in FDD systems, the CSI of the downlink is estimated at the user equipment (UE), then the channel estimate is fed back to the BS through a feedback link with limited bandwidth. However, in massive MIMO systems equipped with a large number of antennas, the required signaling overhead of CSI feedback becomes excessively burdensome.

To address the aforementioned challenges, traditional methods typically adopt codebook-based methods to reduce feedback signaling overhead \cite{codebook1}, such as Type \uppercase\expandafter{\romannumeral1} and Type \uppercase\expandafter{\romannumeral2} codebooks in the 5G New Radio (NR) \cite{3gpp_2}. For instance, a low-complexity codebook search scheme was proposed in \cite{cb3}, which significantly reduces computational complexity. In addition, an effective codebook based on channel statistics was designed to reduce the feedback signaling overhead in \cite{cb1}. Besides, a product codebook was proposed for dual-polarized antenna arrays in \cite{cb2}. However, the complexity of a codebook search algorithm usually increases exponentially with the codebook size. As a result, the overhead grows rapidly, necessitating substantial bandwidth resources in massive MIMO systems. Alternatively, compressed sensing (CS) technologies have been proposed to address the feedback overhead challenge \cite{CS_2}. For example, a CS-based method exploits the CSI sparsity in a specific domain to compress the CSI \cite{CS_0}. Besides, a CS-based method was proposed in \cite{CS_1} that exploits the high spatial correlation of CSI in massive MIMO. However, the CSI recovery of the CS-based methods often requires iterative algorithms, which demand significant computation resources, calling for the development of novel technologies to address these issues.

Recently, deep learning (DL) has experienced significant advancements across various domains, including computer vision \cite{CV}, natural language processing \cite{NLP}, and language models for dialogue \cite{gpt}. Besides, a neural network (NN) offers powerful learning and parallel calculation abilities, making it a promising approach to address the unique challenges of CSI compression. Consequently, researchers have proposed mutiple methods utilizing DL \cite{MIMO_6Gnet}, \cite{6G_intell}. For instance, a DL-based method named CsiNet was proposed in \cite{CSI0}, which adopted the structure of an autoencoder in DL. Specifically, an encoder is deployed at the UE to compress the CSI into a vector, while a corresponding decoder is employed at the BS to recover the original CSI from the compressed vector. Indeed, CsiNet has shown superior accuracy in recovering CSI compared with the codebook-based and CS-based methods.  

To fully harness the potentials of DL, extensive network architectures have been developed to improve the performance of CSI compression and feedback. For example, in \cite{rnn1}, a deep recurrent neural network (RNN) was introduced by considering the time correlation of wireless channels. Besides, a module, termed long short-term memory (LSTM), was introduced to capture the time correlation in \cite{rnn2}. Given that the convolution layer usually plays a crucial role in CSI feedback networks, several researchers have focused on the design of convolution layers to improve the efficiency of the network. For instance, a network named CRNet extracted CSI features at multiple resolutions by incorporating convolution kernels of different sizes \cite{CRNet}. Also, ACRNet, a network introduced in \cite{ACRnet}, aimed to enhance the performance through network aggregation. Besides, a network that can integrate different propagation features was proposed in \cite{DPF}. Furthermore, a self-information model was proposed in \cite{IdasNet} to measure the amount of information in the CSI image and conduct CSI compression by removing information redundancy.

Considering implementation issues in real-world applications, a quantization module was introduced in \cite{csi_quan}, devising a bit-level CSI quantization and feedback network. Moreover, a network added quantization and dequantization modules was proposed in \cite{CSI_FC}. Also considering the presence of noise in the received compressed vectors, \cite{denoise1} proposed a noise-resilient CSI feedback network. In addition, to facilitate network deployment and conserve memory, a lightweight architecture was designed in \cite{ENet} by exploiting the correlations in the angular-delay domain to strike a balance between computational efficiency and performance. Furthermore, another lightweight network was proposed in \cite{CLNet}, which utilized a complex-valued input layer to process signals effectively. Moreover, a distributed machine learning approach for a multiuser mobile edge computing network was proposed in \cite{DML}.

Despite significant research efforts that have been devoted to the exploitation of DL for CSI feedback, existing studies often treat a CSI matrix as an image directly for signal processing \cite{MIMO_edge}, \cite{CSI_MIMO}. However, the characteristics of a CSI matrix are different from those of an image \cite{secure}. In particular, the utilizing of dual-polarized antennas effectively doubles the number of available antennas within a limited antenna-array size, providing a doubling of the spatial multiplexing freedoms. Considering the fact that dual-polarized antennas are widely applied in practice \cite{DP}, the received signals in different polarization directions exhibit similarities from various perspectives, e.g., multipath propagation delay, scattering, angle of arrival (AOA), direction of arrival (DOA), etc. In fact, dual-polarized CSI in different polarization directions tends to demonstrate high correlations within specific sets of factors. To leverage this valuable knowledge as a piece of priori information for NN design, this paper focuses on the characteristics of dual-polarized CSI and develops an enhanced network by leveraging the power of disentangled representation learning. 

Disentangled representation learning is a celebrated technology that can decompose data into independent distinct representations through mutual information (MI) constraints \cite{drl1}. In particular, disentangled representation learning is usually adopted to extract the shared and specific representations from two correlated data sets, where each representation includes a certain physical and semantic significance \cite{DR_RF}.  In the case of dual-polarized CSI, where there is a high correlation between the vertical and horizontal polarizations, we design an encoder network to decompose the original dual-polarized CSI into three parts: the polarization-shared information of the CSI in both polarization directions, the polarization-specific information in the vertical polarization, and the polarization-specific information in the horizontal polarization. Correspondingly, the decoder network recovers the CSI in each polarization by incorporating the polarization-shared and corresponding polarization-specific information.

In this paper, considering the inherent correlation of dual-polarized channels in practice, we propose a deep CSI compression network architecture for handling dual-polarized CSI, named DiReNet, which can differentiate and extract the representations of the polarization-shared and polarization-specific information. In particular, the proposed DiReNet minimizes the information redundancy caused by the polarization correlations. In addition, we establish the network architecture based on the convolutional attention mechanism, which is able to focus its limited attention on crucial information and extract relevant features more effectively \cite{segnext}. The main contributions in this paper are summarized as follows.

\begin{itemize}
\item[$\bullet$] By considering the inherent embedding characteristics of dual-polarized CSI, we propose a network that enables the differentiation of dual-polarized CSI into mutually independent components: the polarization-shared information and two pieces of polarization-specific information. The CSI compression efficiency is improved by minimizing the information redundancy through the disentangled representation learning.
\item[$\bullet$] To manage the network complexity, we design separate fully-connected (FC) networks. Compared with existing networks, the architecture of independent NN reduces the number of encoder parameters by nearly half and that of the decoder by almost one third. The total number of parameters required for the deployment with different compression ratios is significantly reduced.
\item[$\bullet$] To extract the CSI feature maps more effectively, the proposed DiReNet incorporates a convolutional attention mechanism, which combines spatial and channel attentions. Due to the inclusion of attention mechanism, DiReNet selectively focuses its attention on the most essential information, aiming for accurate CSI recovery.
\item[$\bullet$] Experimental results validate the superiority of the proposed DiReNet over other DL-based networks. The proposed method achieves a performance gain of 1.5$\sim$3 dB under different compression ratios, while also reducing the total number of trainable parameters by nearly 1/3 compared to existing state-of-the-art (SOTA) methods.
\end{itemize}

The rest of this paper is organized as follows. The MIMO system model with CSI feedback is introduced in Section \uppercase\expandafter{\romannumeral2}. Section \uppercase\expandafter{\romannumeral3} elaborates on the correlation of dual-polarized CSI and the principle of disentanglement representation learning. Section \uppercase\expandafter{\romannumeral4} describes the details of the proposed DiReNet and the training procedure. Section \uppercase\expandafter{\romannumeral5} presents the simulation results. Conclusions are drawn in Section \uppercase\expandafter{\romannumeral6}.

$\textit{Notations}$: Throughout this paper, plain text denotes scalar variable, boldface lower-case letters denote column vectors, and boldface upper-case letters denote matrices. The superscript $(\cdot)^H$ is the conjugate transpose, the operator $\left\|\cdot\right\|$ is the Euclidean norm, $\mathbb{E}[\cdot]$ is the expectation operator. $\mathbb{R}^{M \times N}$ and $\mathbb{C}^{M \times N}$ represent the real and complex space of $M \times N$ dimensional matrices respectively. $\nabla_{f_\text{NN}}(\mathcal{L})$ represents the gradient of $\mathcal{L}$ with respect to the trainable parameters of an NN $f_\text{NN}$, and $\mathcal{I}(\mathbf{x} ; \mathbf{y})$ denotes the MI between $\mathbf{x}$ and $\mathbf{y}$.

\section{System Model}
We consider the downlink of an FDD massive MIMO system, where the BS equips $N_\text{t}$ dual-polarized antennas and the UE has a single-antenna\footnote{When a UE is equipped with multiple antennas or considering multiuser scenarios, the CSI of each antenna at each user serves as a separate input for training the network.}. The system adopts orthogonal frequency division multiplexing (OFDM) with $N_\text{s}$ subbands. The signal received at the $k$-th subband is expressed as
\begin{equation}
y_k=\mathbf{h}_k^H \mathbf{v}_k x_k+z_k, \quad k=1, \ldots, N_{\text{s}},
\end{equation}
where $\mathbf{h}_k \in \mathbb{C}^{N_\text{t} \times 1}$ and $\mathbf{v}_k \in \mathbb{C}^{N_\text{t} \times 1}$ denote the downlink channel frequency response (CFR) vector at the $k$-th subband and the corresponding precoding vector, respectively. $x_k \in \mathbb{C}$ is the transmitted signal and $z_k \in \mathbb{C}$ represents the additive noise. The downlink CSI in the spatial-frequency domain is written as
\begin{equation}
{\mathbf{H}}=\left[\mathbf{h}_1, \mathbf{h}_2, \ldots,\mathbf{h}_{N_\text{s}}\right]^H,
\end{equation}
where $\mathbf{H} \in \mathbb{C}^{N_\text{s} \times N_\text{t}}$ is the CSI matrix consisting of all the downlink CFR vectors of the $N_\text{s}$ subbands. Since the dual-polarized antenna are widely applied in practice \cite{DP}, we consider that the antennas at the BS are dual-polarized. As such, without loss of generality, the downlink CSI can be divided into two matrices, ${\mathbf{H}}_\text{v} \in \mathbb{C}^{N_\text{s} \times \tfrac{N_\text{t}}{2}}$ and ${\mathbf{H}}_\text{h} \in \mathbb{C}^{N_\text{s} \times \tfrac{N_\text{t}}{2}}$, accounting for the vertical and horizontal polarization directions, respectively, which are expressed by
\begin{subequations}
\begin{align}
{\mathbf{H}}_\text{v}&=\left[\mathbf{h}_{\text{v},1}, \mathbf{h}_{\text{v},2}, \ldots,\mathbf{h}_{\text{v},N_\text{s}}\right]^H,\\
{\mathbf{H}}_\text{h}&=\left[\mathbf{h}_{\text{h},1}, \mathbf{h}_{\text{h},2}, \ldots,\mathbf{h}_{\text{h},N_\text{s}}\right]^H,
\end{align}
\end{subequations}
where $\mathbf{h}_{\text{v},k} \in \mathbb{C}^{\tfrac{N_\text{t}}{2} \times 1}$ and $\mathbf{h}_{\text{h},k} \in \mathbb{C}^{\tfrac{N_\text{t}}{2} \times 1}$ denote the vertical and horizontal polarization CFR vectors at the $k$-th subband, respectively. Since the CSI matrix is complex-valued, we concatenate the real and imaginary parts to form a real-valued matrix as the input.

\begin{figure*}[t]
	\centering
	\includegraphics[width=\linewidth]{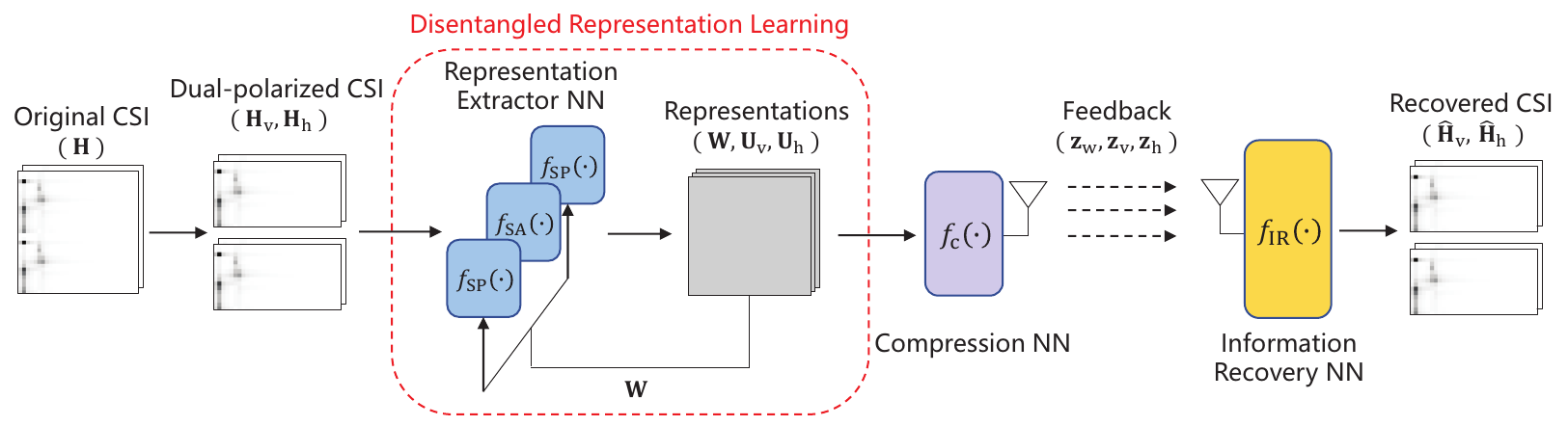}\vspace{-0.3cm}
	\caption{A framework of the proposed DiReNet for dual-polarized CSI.}
	\label{framework}\vspace{-0.2cm}
\end{figure*}

Note that the channel parameters in the vertical and horizontal polarizations generally experience high correlations, as both the polarization magnetic waves propagate along the same physical paths. To exploit the polarization correlations, as depicted in Fig. \ref{framework}, we develop an encoder which not only performs compression, but also disentangles the representations of the two polarizations. For dual-polarized CSI matrices, ${\mathbf{H}_\text{v}}$ and ${\mathbf{H}_\text{h}}$, the encoder extracts the polarization-shared representation, ${\mathbf{W}} \in \mathbb{R}^{2 \times N_\text{s} \times \tfrac{N_\text{t}}{2}}$, and the polarization-specific representations, ${\mathbf{U}_\text{v}} \in \mathbb{R}^{2 \times N_\text{s} \times \tfrac{N_\text{t}}{2}}$ and ${\mathbf{U}_\text{h}} \in \mathbb{R}^{2 \times N_\text{s} \times \tfrac{N_\text{t}}{2}}$, in the vertical and horizontal polarizations, respectively. The obtained representations are then reshaped and compressed into vectors, which are expressed as
\begin{subequations}
\begin{align}
\mathbf{z}_\text{w}&=f_{\text{c}}\left(\mathbf{W}\right)=f_{\text{c}}\left(f_{\text{SA}}\left(\mathbf{H}_\text{v}, \mathbf{H}_\text{h}\right)\right),\\
\mathbf{z}_\text{v/h}&=f_{\text{c}}\left(\mathbf{U}_\text{v/h}\right)=f_{\text{c}}\left(f_{\text{SP}}\left(\mathbf{H}_\text{v/h}, \mathbf{W}\right)\right),
\end{align}
\end{subequations}
where $f_{\text{c}}\left(\cdot\right)$ represents the compression network, $f_{\text{SA}}\left(\cdot\right)$ is the polarization-shared (SA) information extractor network, and $f_{\text{SP}}\left(\cdot\right)$ is the polarization-specific (SP) information extractor network. When the compression ratio is $\sigma > 1$, vectors $\mathbf{z}_\text{w} \in \mathbb{R}^{\frac{2N_\text{s}N_\text{t}}{3\sigma} \times 1}$, $\mathbf{z}_\text{v} \in \mathbb{R}^{\frac{2N_\text{s}N_\text{t}}{3\sigma} \times 1}$, and $\mathbf{z}_\text{h} \in \mathbb{R}^{\frac{2N_\text{s}N_\text{t}}{3\sigma} \times 1}$ denote the compressed polarization-shared vector, the compressed vertical polarization-specific vector, and the compressed horizontal polarization-specific vector, respectively, that to be fed back to the BS.

At the BS side, a corresponding decoder network recovers the CSI in each specific polarization from its compressed polarization-specific vector, i.e., $\mathbf{z}_\text{v}$ or $\mathbf{z}_\text{h}$, and the compressed polarization-shared vector $\mathbf{z}_\text{w}$. It is expressed as
\begin{equation}
\mathbf{\widehat{H}}_\text{v/h}=f_{\text{IR}}\left(\mathbf{z}_\text{v/h}, \mathbf{z}_\text{w}\right),
\end{equation}
where $f_{\text{IR}}\left(\cdot\right)$ is the information recovery (IR) network. $\mathbf{\widehat{H}}_\text{v} \in \mathbb{C}^{N_\text{s} \times \tfrac{N_\text{t}}{2}}$ and $\mathbf{\widehat{H}}_\text{h} \in \mathbb{C}^{N_\text{s} \times \tfrac{N_\text{t}}{2}}$ are the recovered CSI matrix of the vertical and horizontal polarization directions, respectively.

\section{Disentangled Representation Learning for Dual-polarized CSI}

In this section, we first introduce the polarization-shared information and the polarization-specific information within a dual-polarized CSI matrix. Subsequently, we elaborate on the method of disentangled representation learning, which enables the decomposition of dual-polarized CSI into distinct representations of the shared and specific information.

\subsection{Correlation between Dual-polarized CSI}

\begin{figure}[t]
	\centering
	\includegraphics[width=0.65\linewidth]{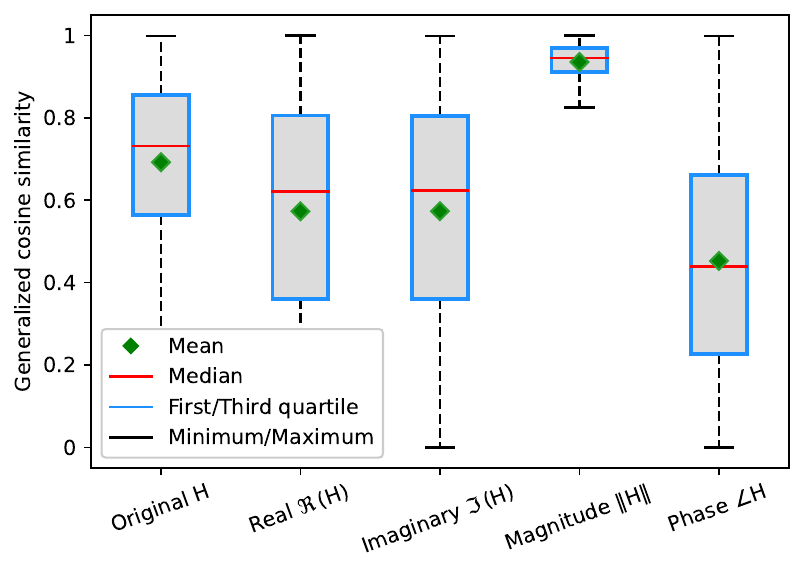}\vspace{-0.5cm}
	\caption{Distribution of the GCS between dual-polarized CSI. (The first quartile is the value below which 25\% of the dataset, The third quartile is the value below which 75\% of the dataset.)}
	\label{gcs}\vspace{-0.3cm}
\end{figure}

We exemplify dual-polarized CSI by utilizing the 3GPP clustered delay line (CDL) channel model \cite{3gpp38901} to demonstrate the correlation between the dual CSI matrices associated with the vertical and horizontal polarizations. The correlation between the two CSI matrices at the two polarization directions is measured using their generalized cosine similarity (GCS) \cite{EVCSI}, which is defined as
\begin{equation}
\rho=\frac{1}{N_\text{s}} \sum_{k=1}^{N_\text{s}} \frac{|{\mathbf{h}^H_{\text{v},k}}{\mathbf{h}_{\text{h},k}}|}{\left\|{\mathbf{h}_{\text{v},k}}\right\|\left\|{\mathbf{h}_{\text{h},k}}\right\|},
\label{eq_six}
\end{equation}
where $\mathbf{h}_{\text{v},k}$ and $\mathbf{h}_{\text{h},k}$ denote the $k$-th subband CSI vectors at the vertical and horizontal polarization directions, respectively.

Note that for the complex-valued CSI, we evaluate both the real and imaginary parts, apart from considering the correlation of the original complex data. To calculate the GCS between the real part of the horizontal CSI and the real part of the vertical CSI, we can also adopt the calculation method in (\ref{eq_six}). Fig. 2 depicts the GCS distribution in the horizontal and vertical polarization directions of the original CSI, the real part of CSI, the imaginary part of CSI, the amplitude of CSI, and the phase of CSI, respectively. We observe that the correlations between the vertical and horizontal polarization CSI of the original data, real part, and imaginary part are rather erratic. Nevertheless, considering the physical nature of the dual-polarized multipath channel model, we recognize that the two different polarization directions can result in random phase differences, while the magnitude should exhibit pronounced correlation due to the shared physical propagation multipaths.

To verify the high correlation in the magnitude domain, we transform the complex-valued CSI into the corresponding polar coordinate representation and separately consider its magnitude and phase correlations. Fig. \ref{gcs} demonstrates that the CSI magnitude in different polarization directions does exhibit a high correlation as expected, whereas the corresponding phases show less correlation. Based on this observation, for dual-polarized CSI, we can conclude that the polarization-shared representation contains highly correlated CSI components such as the magnitude, while the polarization-specific representations include independent information between polarizations, such as random phases.

\subsection{Disentangled Representation Learning in Network Design}

Considering that dual-polarized antennas have been widely applied in practice \cite{DP} and the features of the CSI matrix are different from those of natural images, we propose a model-driven method from a practical perspective. This method aims to reduce information redundancy and enhances the performance by leveraging the inherent polarization correlation of the CSI.

The proposed encoder of the network first disentangles dual-polarized CSI into three independent parts, i.e., a polarization-shared representation, a polarization-specific representation in the vertical polarization, and a polarization-specific representation in the horizontal polarization. Subsequently, it compresses these representations into vectors to effectively reduce the feedback signaling overhead. On the other hand, for the decoder, the proposed network utilizes the received vectors of the polarization-shared $\mathbf{z}_\text{w}$ and vertical polarization-specific vector $\mathbf{z}_\text{v}$ to reconstruct the vertical polarized CSI $\mathbf{H}_\text{v}$. Similarly, the CSI in the horizontal polarization $\mathbf{H}_\text{h}$ is recovered from the polarization-shared vector $\mathbf{z}_\text{w}$ and the horizontal polarization-specific vector $\mathbf{z}_\text{h}$. The proposed architecture consists of three main NN modules, i.e., SA module, SP module, and IR module. These three types of modules are described below.

\begin{figure*}[t]
	\centering
	\includegraphics[width=\linewidth]{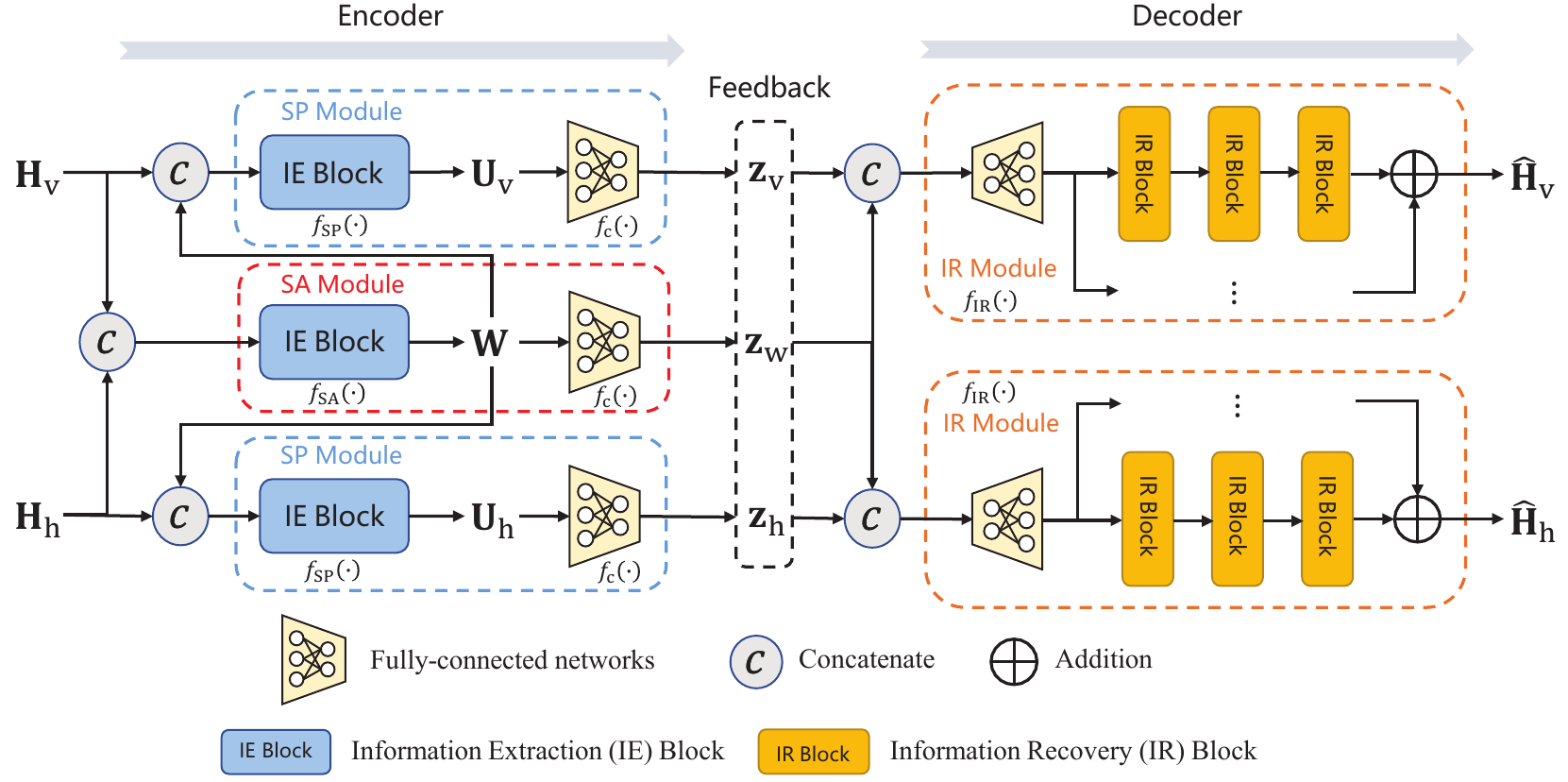}\vspace{-0.3cm}
	\caption{Network architecture of the proposed DiReNet.}
	\label{architecture}\vspace{-0.5cm}
\end{figure*}

$\textit{SA module:}$ This module, framed by the red dashed line in Fig. \ref{architecture}, takes the concatenated vertical and horizontal polarized CSI, i.e., $\mathbf{H}_\text{v}$ and $\mathbf{H}_\text{h}$, as an input. It then captures the information that is shared between the dual polarization directions and obtains the polarization-shared representation $\mathbf{W}$. Finally, this module outputs the compressed polarization-shared vector $\mathbf{z}_\text{w}$. 

$\textit{SP module:}$ This module, framed by the blue dashed lines in Fig. \ref{architecture}, takes the polarization-shared representation $\mathbf{W}$ and the CSI in a certain polarization direction, i.e., either $\mathbf{H}_\text{v}$ or $\mathbf{H}_\text{h}$, as an input. Subsequently, it extracts the exclusive polarization-specific representation, i.e., $\mathbf{U}_\text{v}$ or $\mathbf{U}_\text{h}$, which contains the remaining information excluding the shared information in each polarization direction. In the end, it compresses the representation into a vector, i.e., $\mathbf{z}_\text{v}$ or $\mathbf{z}_\text{h}$.

$\textit{IR module:}$ This module, framed by the orange dashed lines in Fig. \ref{architecture}, reconstructs the CSI in each polarization direction from the polarization-shared information and the corresponding polarization-specific information at the decoder.

As shown in Fig. \ref{architecture}, with the connection structures of the three types of modules described above, the proposed network effectively achieves the disentangled representation learning of dual-polarized CSI \cite{wyner}. In this case, the goal of disentangled representation learning is to extract the shared and specific information between the vertical and horizontal polarizations. In terms of the value of information, we expect the shared representation to contain the common information $\mathcal{I}(\mathbf{H}_{\text{v}}; \mathbf{H}_{\text{h}})$, and the specific representations to contain the conditional entropy $\mathcal{H}(\mathbf{H}_{\text{v}}|\mathbf{H}_{\text{h}})$ or $\mathcal{H}(\mathbf{H}_{\text{h}}|\mathbf{H}_{\text{v}})$, respectively. Based on the structure of the disentangled representation learning network, as the training progresses, the shared representation will contain information in both $\textbf{H}_\text{v}$ and $\textbf{H}_\text{h}$, and the specific representations will contain information in $\textbf{H}_\text{v}$ or $\textbf{H}_\text{h}$, respectively. However, there are two extremes to consider:

$\textit{1)}$ One extreme is that the shared representation does not effectively extract the common information of $\textbf{H}_\text{v}$ and $\textbf{H}_\text{h}$, and the specific representations contain some common information. In other words, the information redundancy cannot be effectively removed.

$\textit{2)}$ The other extreme is that the shared representation contains an excessive amount of information such that it contains non-common information. The network fails to achieve ideal disentanglement.

To avoid these extremes, an MI constraint is imposed, which is mathematically expressed as
\begin{equation}
\mathop{\arg\min}\limits_{\mathbf{W}} \quad \left| \mathcal{I}\left(\mathbf{H}_{\text{v}}, \mathbf{H}_{\text{h}}; \mathbf{W}\right)-\mathcal{I}\left(\mathbf{H}_{\text{v}}; \mathbf{H}_{\text{h}}\right) \right|^2,
\label{gs_seven}
\end{equation}
where $\mathcal{I}\left(\mathbf{H}_{\text{v}}, \mathbf{H}_{\text{h}}; \mathbf{W}\right)$ denotes the MI between ($\mathbf{H}_{\text{v}}, \mathbf{H}_{\text{h}}$) and $\mathbf{W}$. Here, we denote $\mathbf{W}$ as the shared representation of dual-polarized CSI. Then, the MI $\mathcal{I}\left(\mathbf{H}_{\text{v}}, \mathbf{H}_{\text{h}}; \mathbf{W}\right)$ can be viewed as a measure of the complexity of $\mathbf{W}$ \cite{wyner}. Using this approach, we ensure that the network learns to effectively balance between extracting common information and maintaining the uniqueness of specific information, thereby achieving successful disentanglement.

By applying the proposed modules and network architecture, DiReNet performs the representation extraction, compression, and recovery of dual-polarized CSI by fully leveraging the inherent nature of MIMO channel polarizations. Through disentangled representation extraction, the shared information with high correlation and the specific information in each polarization are differentiated and extracted, and the information redundancy in dual-polarized CSI is effectively reduced.

\begin{figure*}[t]
	\centering
	\includegraphics[width=\linewidth]{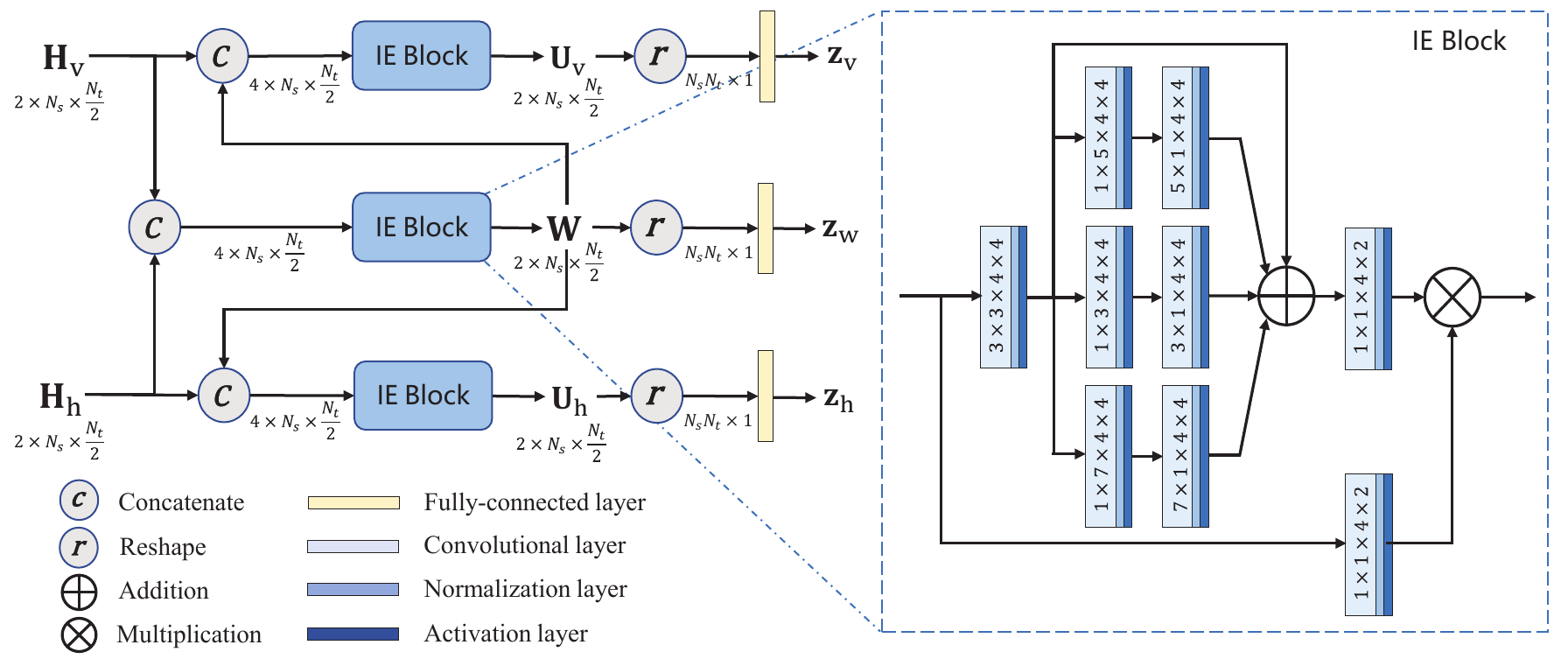}\vspace{-0.3cm}
	\caption{The proposed disentangled learning structure of the encoder.}
	\label{structure_en}\vspace{-0.3cm}
\end{figure*}

\section{Proposed Architecture of DiReNet}
In this section, we first describe the structure details of the encoder and decoder of the proposed DiReNet, which include the design of the convolutional attention block, the separate FC network, and the depth-wise and width-wise extensions. Subsequently, we introduce an NN-based method for the MI estimation. Finally, we describe the training procedure and quantization scheme of the proposed DiReNet.

\subsection{Design of the Encoder Network}
For dual-polarized CSI, we divide it into two matrices, $\mathbf{H}_\text{v}$ and $\mathbf{H}_\text{h}$, corresponding to the vertical and horizontal polarizations, respectively. Since CSI matrices are complex-valued, we concatenate the real and imaginary parts into the corresponding real-valued matrices serving as inputs to the encoder. The encoder contains three components: one SA module and two SP modules. The detailed structure of the encoder is shown in Fig. \ref{structure_en}, where the SA and SP modules share the same structure, consisting of a composite block, named information extraction (IE) block, and an FC layer.

Specifically, the IE block is responsible for extracting the disentangled representations, $\mathbf{W}$, $\mathbf{U}_\text{v}$, and $\mathbf{U}_\text{h}$ from the dual-polarized CSI, while the compressed vectors $\mathbf{z}_\text{w}$, $\mathbf{z}_\text{v}$, and $\mathbf{z}_\text{h}$ are obtained by FC layers. The IE block in Fig. \ref{structure_en} contains three parts: a composite convolution, multiple parallel composite convolution branches, and two $1 \times 1$ composite convolutions. Furthermore, the composite convolution consists of a convolution layer, a normalization layer, and an activation layer. In particular, the normalization layer adopts batch normalization (BN) to stabilize and accelerate the required training. Also, the leakyReLU activation function, with a negative input slope of 0.3, is utilized as a nonlinear transformation. This specific slope has been shown to yield better performance in \cite{CRNet}. Mathematically, the IE block can be expressed as
\begin{subequations}
\begin{align}
\mathbf{M}&=\mathrm{Conv}_{1 \times 1}\left(\sum_{i=0}^3 \mathrm{B}_i\left(\mathrm{Conv}_{\text{m} \times \text{n}}\left(\mathbf{H}_\text{in}\right)\right)\right),  \\
\mathbf{H}_\text{out}&=\mathrm{Conv}_{1 \times 1}\left(\mathbf{H}_\text{in}\right) \otimes \mathbf{M},
\end{align}
\end{subequations}
where $\mathbf{M} \in \mathbb{R}^{2 \times N_\text{s} \times \tfrac{N_\text{t}}{2}}$ represents the attention mask, $\mathbf{H}_\text{in} \in \mathbb{R}^{4 \times N_\text{s} \times \tfrac{N_\text{t}}{2}}$ and $\mathbf{H}_\text{out} \in \mathbb{R}^{2 \times N_\text{s} \times \tfrac{N_\text{t}}{2}}$ are the input and output features respectively, $\otimes$ is the matrix multiplication operator, $\mathrm{B}_i$, $i \in\{0,1,2,3\}$ denotes the $i$-th branch, and $\mathrm{Conv}_{\text{m} \times \text{n}}$ denotes the composite convolution with a kernel size of $m \times n$.

As shown in Fig. \ref{structure_en}, the composite convolution extracts an initial CSI feature map from the input, which is concatenated by two matrices. The SA module is expected to extract the feature map containing the highly correlated channel information between $\mathbf{H}_\text{v}$ and $\mathbf{H}_\text{h}$. As for the SP module, the feature map is intended to capture the remaining information in $\mathbf{H}_\text{v}$ or $\mathbf{H}_\text{v}$, excluding the polarization-shared representation $\mathbf{W}$.

The multiple parallel composite convolution branches are designed to capture the spatial correlation at different scales, since the sparsity of the CSI varies across different channel scenarios. Therefore, the kernel size for each branch is set to 3, 5, and 7, respectively. Besides, two composite convolutions are employed in all branches, as shown in Fig. 4.

The $1 \times 1$ composite convolutions integrate features from different channels, and there are two $1 \times 1$ convolution layers: one outputting a processed input and the other outputting an attention mask. The final output is obtained by multiplying the processed input with the attention mask. Thanks to the involved attention mechanism, the IE block can extract information selectively by focusing limited attention on the essential information for improving the CSI recovery accuracy. 

The FC network is designed to obtain the compressed vectors. We design the separate FC network that is different from the ones in existing frameworks, e.g., \cite{CSI0}, \cite{CRNet}, \cite{ACRnet}, to ensure independence between different representations and to minimize the parameters within the FC network. In the following, we compare the parameters between these methods. First, for a typical framework, a single FC network is adopted. Given that the size of the CSI feature map is $2 \times N_\text{s} \times N_\text{t}$ and the length of the vector is $2N_\text{s}N_\text{t}$, the number of parameters of the FC network for a typical framework is given by
\begin{equation}
{P}_{0}= 2N_\text{s}N_\text{t} \times \frac{2N_\text{s}N_\text{t}}{\sigma}= \frac{4{N_\text{s}}^2{N_\text{t}}^2}{\sigma}.
\label{para0}
\end{equation}

In contrast, as shown in Fig. \ref{structure_en}, the proposed CSI feedback framework employs the design of three parallel FC networks. Since the three FC networks are designed separately and the length of each input vector to the FC network is ${N_\text{s}N_\text{t}}$, the number of parameters in the FC network can be calculated as
\begin{equation}
{P}_{1}= 3 \times N_\text{s}N_\text{t} \times \frac{2N_\text{s}N_\text{t}}{3\sigma}= \frac{1}{2}{P}_{0}.
\label{para1}
\end{equation}

By comparing the numbers of parameters in (\ref{para0}) and (\ref{para1}), it is concluded that the FC network in the encoder reduces the number of trained parameters by half compared to the typical existing networks. In addition, the design of separate FC networks enables independent compression for better capturing the unique characteristics of the polarization-shared and polarization-specific representations.

\subsection{Design of the Decoder Network}
\begin{figure*}[t]
	\centering
	\includegraphics[width=\linewidth]{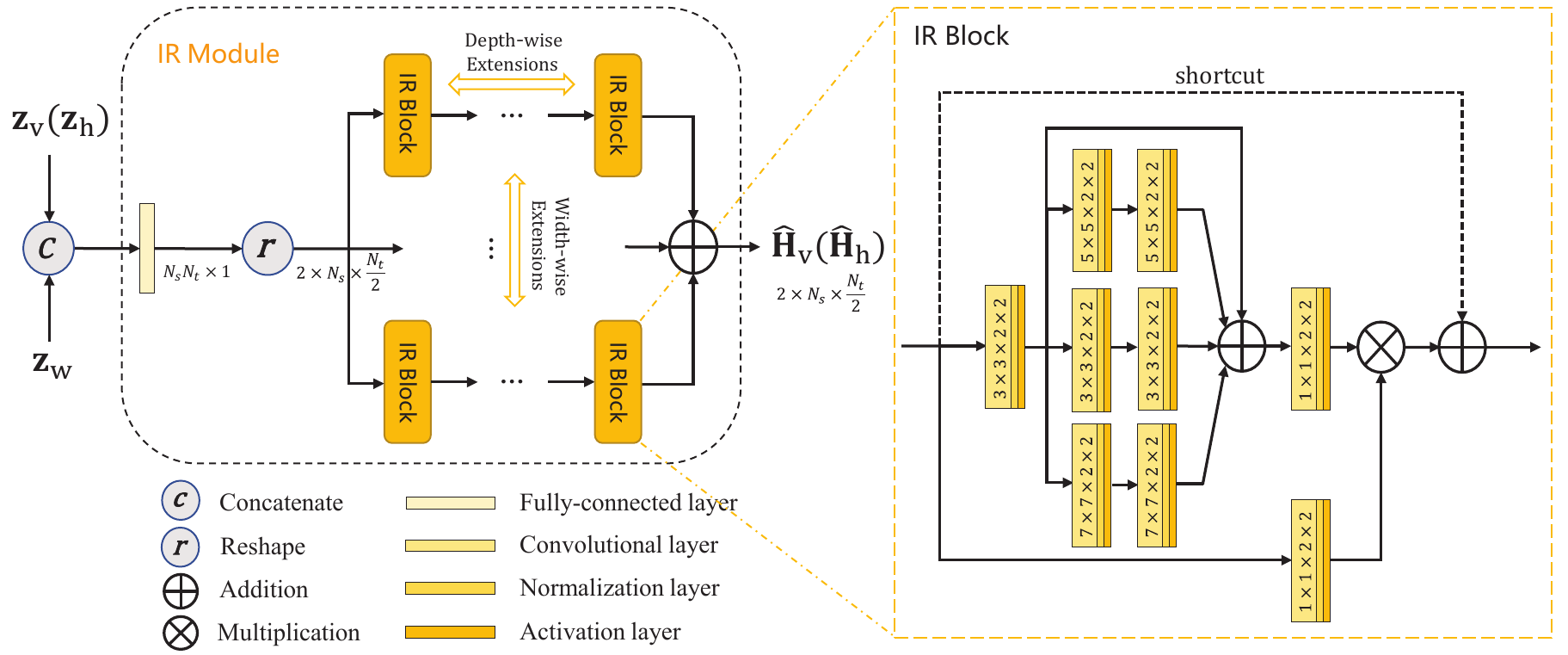}\vspace{-0.3cm}
	\caption{The structure of the IR module.}
	\label{structure_de}\vspace{-0.3cm}
\end{figure*}

The decoder consists of two IR modules for recovering CSI in respective polarization directions from vectors. Fig. \ref{structure_de} illustrates the detailed structure of the IR module. The IR module first restores the dimension by FC networks and reshapes it into a CSI feature map. Then, the CSI is reconstructed by combining the shared information and the specific information corresponding to the polarization direction.

For typical decoders in the literature, the adopted FC networks are symmetric with the encoder. As such, the number of parameters in these FC networks can also be expressed as in (\ref{para0}). By contrast, for the proposed DiReNet decoder, the design of separate FC networks is shown in Fig. \ref{structure_de} and the required number of parameters is
\begin{equation}
{P}_{2}= 2 \times N_\text{s}N_\text{t} \times \left( \frac{2N_\text{s}N_\text{t}}{3\sigma} + \frac{2N_\text{s}N_\text{t}}{3\sigma} \right)= \frac{2}{3} {P}_{0}.
\label{para2}
\end{equation}

By comparing the numbers of parameters in (\ref{para0}) and (\ref{para2}), we observe that the number of parameters in the FC network is reduced by one third compared with the typical frameworks. In the proposed DiReNet decoder, the reshaped CSI feature map is fed into the network with the extensions of the IR block to recover CSI. In particular, there are two types of network extensions: depth-wise and width-wise. 

On one hand, the depth-wise extension introduces extra IR blocks in series resulting in a deeper network. It has been reported in \cite{depth1} that deeper models generally perform better in learning complex transformations and complicated functions than a compact shallow architecture, due to superior nonlinear representation capabilities of the former.

On the other hand, the width-wise extension adds IR blocks in parallel. It has been revealed that as the width of the network increases, the extracted features containing spatial information becomes richer \cite{width1}. Yet, the application of width-wise extension also increases the network complexity. To strike a balance of complexity and performance, 3 IR blocks are connected in series, forming a path for depth-wise extension. Meanwhile, 5 paths are connected in parallel, constituting the main part of the IR module in the width-wise extension.

The design of the IR block closely resembles that of the IE block. However, considering that the BS is equipped with more memory and computing power, the IR block replaces the convolution kernels in the IE block with larger convolution kernels, e.g., a pair of 3$\times$3 and 3$\times$3 kernels instead of a pair of 1$\times$3 and 3$\times$1 kernels, the kernel size for each branch is also set to 3, 5, and 7, respectively. As shown in Fig. \ref{structure_de}, to prevent gradient exploding and vanishing during training, the IR block incorporates a shortcut connection.

\subsection{The Mutual Information Estimator}
To extract a sufficient and effective representation as in (\ref{gs_seven}), the MI must be obtained accurately. However, directly calculating the MI in a high dimensional space is generally intractable. Since an NN enables to approximate arbitrary functions, estimating MI via DL-based methods has become a widespread approach. We apply the contrastive log-ratio upper bound (CLUB) of MI in \cite{club} for the optimization of estimating the MI. The CLUB estimator implemented by an NN effectively reduces the computational complexity and demonstrates attractive performance for MI estimation tasks. For a pair of correlated data $\mathbf{x}$ and $\mathbf{y}$, the MI is expressed as
\begin{equation}
\mathcal{I}\left(\mathbf{x};\mathbf{y}\right)=\mathbb{E}_{p\left(\mathbf{x},\mathbf{y}\right)}\left[\ln p\left(\mathbf{y}|\mathbf{x}\right)\right]-\mathbb{E}_{p\left(\mathbf{x}\right)} \mathbb{E}_{p\left(\mathbf{y}\right)}\left[\ln p\left(\mathbf{y}|\mathbf{x}\right)\right].
\label{gs_club}
\end{equation}

In practice, the distribution $p\left(\mathbf{y}|\mathbf{x}\right)$ is typically implemented using NN and the estimated method can be easily extended to tensors. We adopt the network $f_{\text{MI}}\left(\cdot\right)$ to approximate the actual distribution $p\left(\mathbf{y}|\mathbf{x}\right)$. With the powerful learning ability of NN, the MI is estimated effectively by the CLUB estimator, and it has been shown in \cite{club} that the estimator converges and the estimated MI approaches the actual MI by maximizing the log-likelihood of $p(\mathbf{y}|\mathbf{x})$. For data samples $\left\{\left(\mathbf{x}_i, \mathbf{y}_i\right)\right\}_{i=1}^N$, the log-likelihood function, that is the loss function used to update the estimator $f_{\text{MI}}\left(\cdot\right)$, is expressed as
\begin{equation}
\mathcal{L}_{\text{CLUB}}=\frac{1}{N} \sum_{i=1}^N \ln p\left(\mathbf{y}_i|\mathbf{x}_i\right)=\frac{1}{N} \sum_{i=1}^N \ln (f_{\text{MI}}\left(\mathbf{x}_i,\mathbf{y}_i\right)).
\label{gs_loss_club}
\end{equation}

\subsection{Training Procedure of DiReNet}
We now introduce the design of the training procedure that incorporates MI regularization. Due to the fact that online training requires continuously broadcasting a large number of weights, resulting in a significant cost of communication resources, the proposed DiReNet is trained offline. This means that the parameters of the encoder and decoder are static on the UE and BS, respectively. For explanatory purposes, we denote the parameters of DiReNet as $\mathbf{\Phi}=\left\{ \mathbf{\Phi}_\text{EN}, \mathbf{\Phi}_\text{DE}\right\}$, where $\mathbf{\Phi}_\text{EN}$ is the parameters of encoder, $\boldsymbol{\Phi}_\text{DE}$ is the parameters of decoder. The reconstructed CSI is expressed by
\begin{equation}
\mathbf{\widehat{H}}_{\text{v}/\text{h}} = f_{\text{DE}}\left(f_{\text{EN}}\left(\mathbf{H}_{\text{v}/\text{h}} ; \mathbf{\Phi}_\text{EN}\right) ; \mathbf{\Phi}_\text{DE} \right),
\end{equation}
where $f_{\text{EN}}\left(\cdot\right)$ denotes the encoder network and $f_{\text{DE}}\left(\cdot\right)$ denotes the decoder network. Note that the input and output of DiReNet are normalized such that the scaled elements are in the range of [0, 1]. Besides, the mean squared error (MSE) is exploited as the loss function, which is given as
\begin{equation}
\mathcal{L}_{\text{MSE}}=\frac{1}{2T} \sum_{t=1}^T \left( \left\| \mathbf{\widehat{H}}_{\text{v},t} - \mathbf{H}_{\text{v},t}\right\|^2 + \left\| \mathbf{\widehat{H}}_{\text{h},t} - \mathbf{H}_{\text{h},t}\right\|^2 \right),
\label{gs_loss_mse}
\end{equation}
where $T$ is the total number of samples in the training set, $\mathbf{H}_{\text{v},t}$ and $\mathbf{H}_{\text{v},t}$ denote the original CSI of the $t$-th sample on the vertical and horizontal polarizations, respectively, and $\mathbf{\widehat{H}}_{\text{v},t}$ and $\mathbf{\widehat{H}}_{\text{h},t}$ denote the reconstructed CSI of $t$-th sample on the vertical and horizontal polarizations, respectively.

In DL frameworks, especially in the representation learning, MI is commonly used as a regularizer in loss functions, to encourage or restrict the dependence between different data. By incorporating different regularizers, we can enhance the disentanglement performance or facilitate the interpretability of the learned representation. The MI constraint in (\ref{gs_seven}) implies that the ideal shared representation $\textbf{W}$, which includes common information, is obtained by minimizing the difference between the mutual information of the original CSI with shared representation and the mutual information of CSIs on different polarization directions. For disentanglement achieved through neural networks, $\textbf{W}$ is the output of the encoder of the neural network, and it can be altered by adjusting the network parameters. Therefore, (\ref{gs_seven}) is added as a loss function for training the entire neural network and obtaining the ideal disentanglement representation, which is expressed as
\begin{equation}
\mathcal{L}_{\text{MI}}=\frac{1}{T} \sum_{t=1}^T  \left| \mathcal{I}\left(\mathbf{H}_{\text{v},t}, \mathbf{H}_{\text{h},t}; \mathbf{W}_t\right)-\mathcal{I}\left(\mathbf{H}_{\text{v},t}; \mathbf{H}_{\text{h},t}\right) \right|^2,
\label{gs_loss_mi}
\end{equation}
where the MI is estimated by the CLUB estimator defined in (\ref{gs_club}) and the NN is trained by the loss function in (\ref{gs_loss_club}). The results of $\mathbf{W}_t$ is the polarization-shared representation of the $t$-th sample, $\mathcal{I}\left(\mathbf{H}_{\text{v},t}, \mathbf{H}_{\text{h},t}; \mathbf{W}_t\right)$ denotes the complexity of the polarization-shared representation, and $\mathcal{I}\left(\mathbf{H}_{\text{v},t}; \mathbf{H}_{\text{h},t}\right)$ is the MI between the correlated dual-polarized CSI.

By setting the difference between these two MI functions as the optimization objective, we expect to obtain sufficient and effective representations. In summary, combining the above two loss functions in (\ref{gs_loss_mse}) and (\ref{gs_loss_mi}), we set $\lambda$ as a hyper-parameter to balance the magnitudes of $\mathcal{L}_\text{MSE}$ and $\mathcal{L}_\text{MI}$. The adopted loss function is written as
\begin{equation}
\mathcal{L} = \mathcal{L}_{\text{MSE}} + \lambda \mathcal{L}_{\text{MI}}.
\label{gs_loss_add}
\end{equation}

\begin{algorithm}[t]
\caption{Training Procedure of Proposed DiReNet}
\label{alg1}
\begin{algorithmic}
\STATE 
\textbf{Input:} Training dataset $\mathcal{T}$, batch size $B$. \\
\textbf{Output:} $f_{\text{EN}}^*$, $f_{\text{DE}}^*$, $f_{\text{MI}_{1}}^*$, and $f_{\text{MI}_{2}}^*$.  \\
\textbf{Parameters:} Learning rate $\eta$, coefficients $\lambda$.
\REPEAT 
\STATE
Draw batch data $\left(\mathbf{H}_{\text{v},t}, \mathbf{H}_{\text{h},t}\right)$ from $\mathcal{T}$;
\STATE
\textbf{Step 1:} \\
Compute $\mathbf{\widehat{H}}_{\text{v},t}, \mathbf{\widehat{H}}_{\text{h},t}=f_{\text{DE}}\left(f_{\text{EN}}\left(\mathbf{H}_{\text{v},t}, \mathbf{H}_{\text{h},t}\right)\right)$;\\
Obtain $\mathbf{W}_{t}$ from the forward process;\\
Compute $\mathcal{I}\left(\mathbf{H}_{\text{v},t}, \mathbf{H}_{\text{h},t}; \mathbf{W}_t\right)$ and $\mathcal{I}\left(\mathbf{H}_{\text{v},t}; \mathbf{H}_{\text{h},t}\right)$ according to (\ref{gs_club});\\
Compute $\mathcal{L} = \mathcal{L}_{\text{MSE}} + \lambda \mathcal{L}_{\text{MI}}$ according to (\ref{gs_loss_mse})-(\ref{gs_loss_add});\\
Update $f_{\text{EN}} \leftarrow f_{\text{EN}}-\eta \nabla_{f_{\text{EN}}}(\mathcal{L}_{\text{MSE}} + \lambda \mathcal{L}_{\text{MI}}), f_{\text{DE}} \leftarrow f_{\text{DE}}-\eta \nabla_{f_{\text{DE}}}(\mathcal{L}_{\text{MSE}} + \lambda \mathcal{L}_{\text{MI}})$;\\
\textbf{Step 2:} \\
Compute $\mathcal{L}_{\text{CLUB}}$ according to (\ref{gs_loss_club});\\
Update $f_{\text{MI}_{1}} \leftarrow f_{\text{MI}_{1}}-\eta \nabla_{f_{\text{MI}_{1}}}(\mathcal{L}_{\text{CLUB}}), f_{\text{MI}_{2}} \leftarrow f_{\text{MI}_{2}}-\eta \nabla_{f_{\text{MI}_{2}}}(\mathcal{L}_{\text{CLUB}})$;\\
\UNTIL{convergence}
\RETURN{$f_{\text{EN}}$, $f_{\text{DE}}$, $f_{\text{MI}_{1}}$, and $f_{\text{MI}_{2}}$.}
\end{algorithmic}
\end{algorithm}

The loss function consists of the NMSE term for measuring CSI recovery accuracy and the MI constraint term used to extract disentangled representations. Even if the dual polarization correlation of the input CSI is low or absent, the proposed network can still adaptively extract the features of CSI through the NMSE loss function for compression and recovery.

Since the input dimensions of the two MI in (\ref{gs_loss_mi}) are different, we construct two different estimators $f_{\text{MI}_{1}}\left(\cdot\right)$ and $f_{\text{MI}_{2}}\left(\cdot\right)$ in the training procedure. Specifically, $f_{\text{MI}_{1}}\left(\cdot\right)$ is adopted to estimate the MI between polarization-shared representation $ \mathbf{W}_t$ and original dual-polarized CSI ($\mathbf{H}_{\text{v},t}, \mathbf{H}_{\text{h},t}$), $f_{\text{MI}_{2}}\left(\cdot\right)$ is adopted for the MI estimation between the vertical polarization CSI $\mathbf{H}_{\text{v},t}$ and the horizontal polarization CSI $\mathbf{H}_{\text{h},t}$. In addition, since the estimated MI is required in the training process of the encoder and decoder, the training procedure of the proposed DiReNet is achieved through alternating and iterative execution. The training algorithm is described in Algorithm \ref{alg1}, and the training procedure consists of the following two steps. 

$\textit{Step 1:}$ Fixing $f_{\text{MI}_{1}}\left(\cdot\right)$ and $f_{\text{MI}_{2}}\left(\cdot\right)$, we jointly optimize $f_{\text{EN}}\left(\cdot\right)$ and $f_{\text{DE}}\left(\cdot\right)$ to achieve the compression and recovery of dual-polarized CSI. Specifically, we adopt the gradient descent algorithm to update $f_{\text{EN}}\left(\cdot\right)$ and $f_{\text{DE}}\left(\cdot\right)$ that are given by
\begin{subequations}
\begin{align}
f_{\text{EN}} &\leftarrow f_{\text{EN}}-\eta \nabla_{f_{\text{EN}}}(\mathcal{L}_{\text{MSE}} + \lambda \mathcal{L}_{\text{MI}}), \\ 
f_{\text{DE}} &\leftarrow f_{\text{DE}}-\eta \nabla_{f_{\text{DE}}}(\mathcal{L}_{\text{MSE}} + \lambda \mathcal{L}_{\text{MI}}),
\end{align}
\end{subequations}
respectively, where $\eta>0$ is the learning rate and $\lambda>0$ is the hyper-parameter to balance the magnitudes.

$\textit{Step 2:}$ Fixing $f_{\text{EN}}\left(\cdot\right)$ and $f_{\text{DE}}\left(\cdot\right)$, we optimize $f_{\text{MI}_{1}}\left(\cdot\right)$ and $f_{\text{MI}_{2}}\left(\cdot\right)$ to learn the MI estimators, the update process of  $f_{\text{MI}_{1}}\left(\cdot\right)$ and $f_{\text{MI}_{2}}\left(\cdot\right)$ follows
\begin{subequations}
\begin{align}
f_{\text{MI}_{1}} &\leftarrow f_{\text{MI}_{1}}-\eta \nabla_{f_{\text{MI}_{1}}}(\mathcal{L}_{\text{CLUB}}), \\
f_{\text{MI}_{2}} &\leftarrow f_{\text{MI}_{2}}-\eta \nabla_{f_{\text{MI}_{2}}}(\mathcal{L}_{\text{CLUB}}).
\end{align}
\end{subequations}

As the learning progresses, the estimators $f_{\text{MI}_{1}}\left(\cdot\right)$ and $f_{\text{MI}_{2}}\left(\cdot\right)$ are gradually trained to accurately estimate the MI. Consequently, accurate MI estimation is beneficial for the encoder $f_{\text{EN}}\left(\cdot\right)$ and decoder $f_{\text{DE}}\left(\cdot\right)$ to effectively realize the disentangled representations extraction, compression, and recovery of the dual-polarized CSI.

In practice, the transmission of continuous-valued vectors is difficult, therefore it is necessary to quantize the continuous-valued vectors. The encoder of the proposed DiReNet obtains vectors with different semantics, namely a polarization-shared vector and two polarization-specific vectors in the vertical and horizontal polarizations. For these vectors, distinct quantization-levels are adopted to achieve flexible feedback accuracy. 

\begin{figure*}[t]
	\centering
	\includegraphics[width=\linewidth]{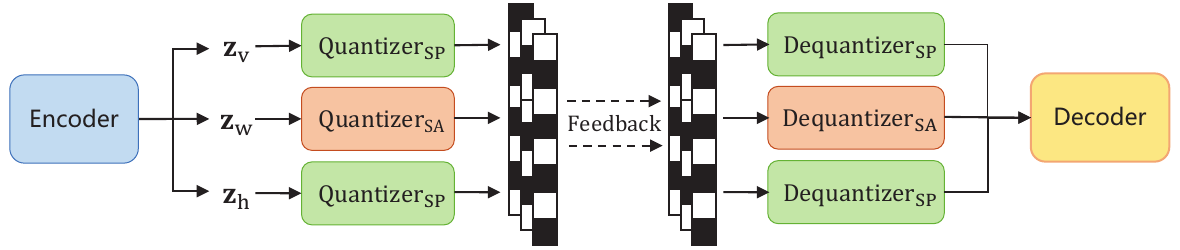}\vspace{-0.3cm}
	\caption{The proposed DiReNet with CSI quantization.}
	\label{quantizer}\vspace{-0.2cm}
\end{figure*}

As shown in Fig. \ref{quantizer}, independent quantizers are designed to quantize the vectors with different quantization-levels. In the decoder, corresponding dequantizers are exploited to reconstruct the continuous-valued vectors. Specifically, the quantization is achieved by the non-learning uniform quantizers and dequantizers. The flexible feedback accuracy is achieved by setting distinct quantization-levels. In particular, during the training phase, we temporarily exclude the quantization operation. While in the deployment phase, the vectors obtained by the encoder are quantized through the quantizers and then fed back to the decoder. Correspondingly, dequantizers are exploited to recover the continuous-valued vectors in the decoder.

\section{Experimental Results}
In this section, we evaluate the effectiveness of our proposed DiReNet for CSI compression and recovery. First, we describe the details of our experiments, including data sets, evaluation criteria, parameter settings, and hardware platform. Then, we show the performance comparison between the proposed DiReNet and existing DL-based methods under different compression ratios, and we also discuss the network performance with quantization feedback. Subsequently, we present the effectiveness of disentangled representation learning and the MI regularization in the proposed framework. Then, we discuss the impact of depth-wise and width-wise extensions in the decoder. Finally, we discuss the generalization abilities and the achievable rate performance of the proposed DiReNet.

\subsection{Simulation Setup}
In our experiments, we generate the training set, validation set, and testing set through CDL channel models defined in 3GPP specification TR 38.901 for link-level simulations \cite{3gpp38901}. Table \ref{sim_para} list the basic parameters, we utilize three channel models with different system configurations for simulations, including the CDL-A channel model with a delay spread of 30 ns and a UE speed of 3 km/h, the CDL-B channel model with a delay spread of 100 ns and a UE speed of 30 km/h, and the CDL-C channel model with a delay spread of 300 ns and a UE speed of 120 km/h. The CDL-A channel is exploited as the default dataset in most of our subsequent simulations. Besides, we represent the polarization correlation of the CSI by calculating the GCS between the CSI in horizontal and vertical polarization directions. A larger GCS value indicates a higher polarization correlation. We calculate the magnitude GCS values for various channel data and list their means and standard deviations in Table \ref{sim_para}. It is observed that the polarization correlations for the CDL-A, CDL-B, and CDL-C channels decrease in order.

\begin{table}[tbp]
    \center
    \caption{Basic Simulation Parameters}
    \setlength{\tabcolsep}{1.5mm} 
    \renewcommand{\arraystretch}{1}
    \begin{tabular}{c|c|c|c}
    \Xhline{1.1pt} Parameter & \multicolumn{3}{c}{Value} \\
    \hline \hline Carrier frequency & \multicolumn{3}{c}{$3.5\ \mathrm{GHz}$} \\
    \hline Bandwidth & \multicolumn{3}{c}{$10\ \mathrm{MHz}$} \\
    \hline Subcarrier spacing & \multicolumn{3}{c}{$15\ \mathrm{kHz}$} \\
    \hline Subband numbers & \multicolumn{3}{c}{$32$} \\
    \hline $N_\text{t}, N_r$ & \multicolumn{3}{c}{$32, 1$} \\
    \hline Channel model \cite{3gpp38901} & CDL-A & CDL-B & CDL-C \\
    \hline Delay spread & $30 \mathrm{~ns}$ & $100 \mathrm{~ns}$ & $300 \mathrm{~ns}$ \\
    \hline UE speed & $3 \mathrm{~km} / \mathrm{h}$ & $30 \mathrm{~km} / \mathrm{h}$ & $120 \mathrm{~km} / \mathrm{h}$ \\
    \hline The value of GCS & $0.936 \pm 0.047$ & $0.869 \pm 0.065$ & $0.741 \pm 0.092$ \\
    \hline Channel estimation & \multicolumn{3}{c}{Ideal} \\
    \hline$M_\text{train}, M_\text{val}, M_\text{test}$ & \multicolumn{3}{c}{$1000, 200, 200$} \\
    \hline$T_\text{train}, T_\text{val}, T_\text{test}$ & \multicolumn{3}{c}{$100, 100, 100$} \\
    \Xhline{1.1pt}
    \end{tabular}
    \label{sim_para}
\end{table}

For each training set, validation set, and testing set, we sample dual-polarized CSI of $M$ users and $T$ time slots, with a total of $M\times{T}$ CSI matrices, where $M_\text{train}$ = 1,000, $M_\text{val}$ = 200, $M_\text{test}$ = 200, and $T_\text{train}$ = $T_\text{val}$ = $T_\text{test}$ = 100, thus the training set, validation set, and testing set contain 100,000, 20,000, and 20,000 samples, respectively. The number of BS dual-polarized antennas is $N_\text{t}$ = 32 and the number of subbands is $N_\text{s}$ = 32, where we choose 12 subcarriers as one resource block (RB) and every two RBs as one subband. For the case of expanding to a different number of subcarriers, altering the number of RBs contained in one subband or training the network with different input dimensions are feasible approaches. As for the depth-wise and width-wise extensions, we set them to 3 and 5, respectively. At the training phase, the hyper-parameter $\lambda$ is set to $\text{10}^{-\text{5}}$, the adaptive momentum (Adam) optimizer \cite{adam} with learning rate $\eta = 0.001$ is adopted, the number of training epochs is 1,000, the batch size is 200, and the simulation is carried out in Pytorch on a GTX3090 GPU. 

\subsection{Performance of the Proposed DiReNet}

To evaluate the performance of the proposed DiReNet for CSI compression and recovery, the compression ratios in our simulation experiments are 8, 16, 32, and 64, respectively. We compare with existing methods in terms of the normalized mean square error (NMSE) performance and the network complexity. The existing DL-based methods proposed by other researchers for comparison include CsiNet \cite{CSI0}, CRNet \cite{CRNet}, and ACRNet \cite{ACRnet}. Among these, ACRNet is a method that enhances performance through network aggregation and supports networks of different scales, such as ACRNet-1x and ACRNet-5x. Given that ACRNet is currently a SOTA method used for CSI feedback, we selected both ACRNet-1x and ACRNet-5x for comparison with our proposed DiReNet. 

\subsubsection{NMSE Performance}
We utilize the NMSE between the original dual-polarized CSI ($\mathbf{H}_\text{v}$, $\mathbf{H}_\text{h}$) and the recovered dual-polarized CSI ($\mathbf{{\widehat{H}}}_\text{v}$, $\mathbf{{\widehat{H}}}_\text{h}$), as a metric to measure CSI recovery accuracy, which is defined as
\begin{equation}
\mathrm{NMSE}=\frac{1}{2}\ \mathbb{E}\left[ \frac{\left\|\mathbf{\widehat{H}}_\text{v}-\mathbf{H}_\text{v}\right\|^2}{\left\|\mathbf{H}_\text{v}\right\|^2}+\frac{\left\|\mathbf{\widehat{H}}_\text{h}-\mathbf{H}_\text{h}\right\|^2}{\left\|\mathbf{H}_\text{h}\right\|^2} \right].
\end{equation}

We evaluate the performance of the proposed DiReNet, without quantization, by the cumulative distribution function (CDF) of the NMSE, which can illustrate the percentage of CSI recovery performance for the entire testing set. By analyzing the CDF of the NMSE, we can observe the stability of the trained model and the performance gain of our proposed DiReNet compared to existing DL-based methods.

\begin{figure*}[t]  
    \begin{center}
        \subfigcapskip=-5pt
        \subfigure[$\sigma$ = 8.]{\includegraphics[width=0.49\linewidth]{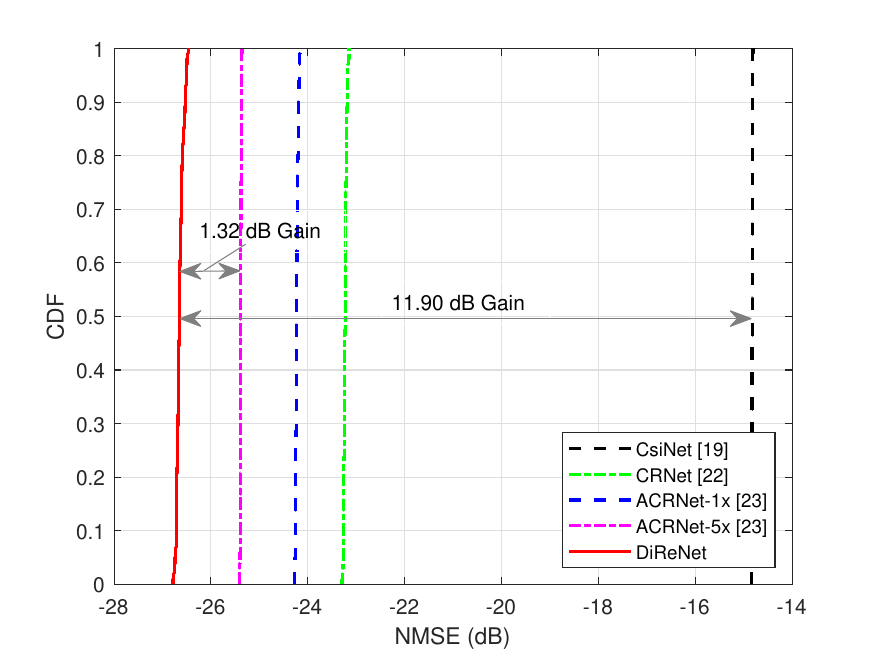}}\hspace{-5pt}
        \subfigure[$\sigma$ = 16.]{\includegraphics[width=0.49\linewidth]{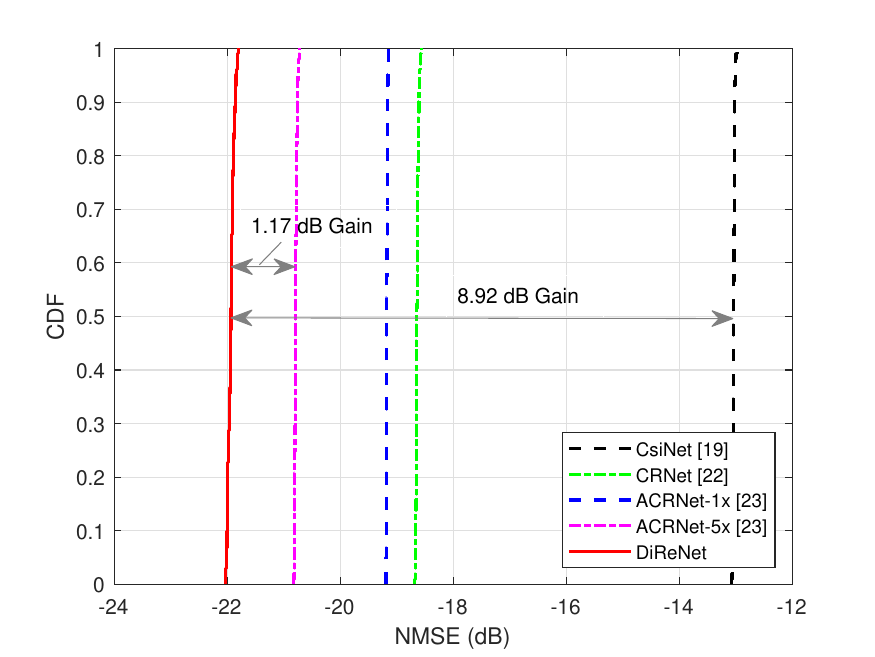}}\vspace{-5pt}
        \subfigure[$\sigma$ = 32.]{\includegraphics[width=0.49\linewidth]{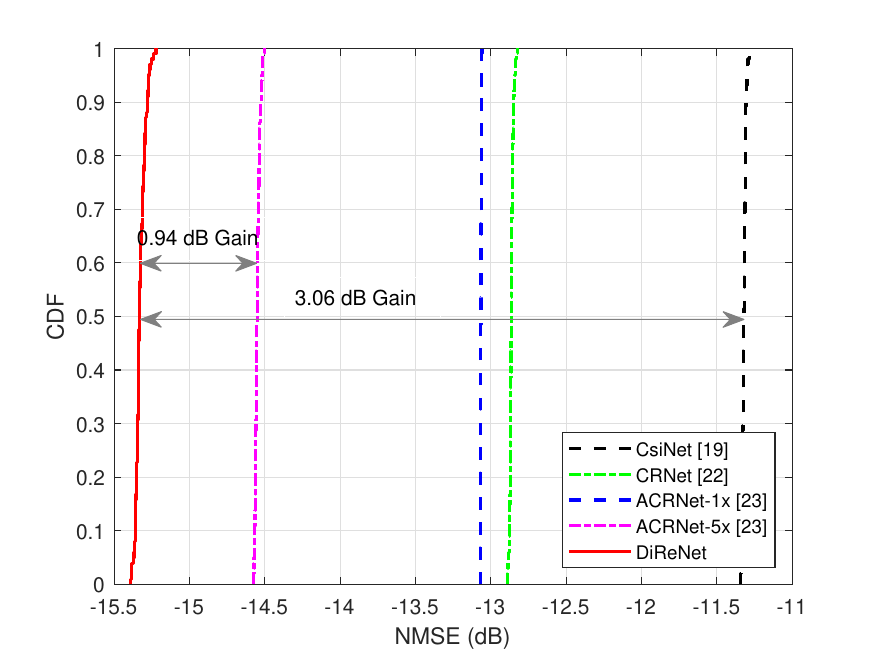}}\hspace{-5pt}
        \subfigure[$\sigma$ = 64.]{\includegraphics[width=0.49\linewidth]{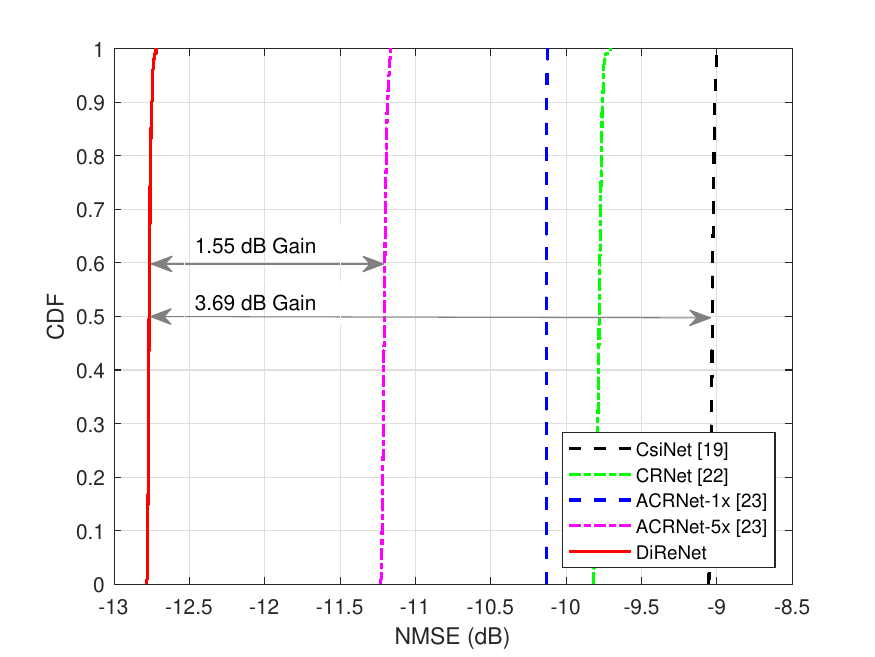}}\vspace{-0.1cm}
        \caption{CDF of NMSE performance without quantization under different compression ratio $\sigma$.}\vspace{-0.25cm}
        \label{cdf}
    \end{center}\vspace{-0.3cm}
\end{figure*}

As shown in Fig. \ref{cdf}, we observe that the proposed DiReNet outperforms other networks under all the considered compression ratios. When the compression ratio $\sigma$ is 8, the best NMSE of DiReNet, CRNet, and ACRNet-1x are -26.74 dB, -23.30 dB, and -24.23 dB, respectively, and the NMSE performance gain by DiReNet is approximately 3 dB. Also, even at a high compression ratio of 64, the NMSE of DiReNet, CRNet, and ACRNet-1x are -12.76 dB, -9.82 dB, and -10.14 dB, respectively, and the NMSE performance gain by the proposed DiReNet is also about 3 dB. Compared to the larger advanced network named ACRNet-5x, DiReNet also has better performance. Specifically, the NMSE of DiReNet and ACRNet-5x are -26.74 dB and -25.42 dB at the compression ratio of 8. When the compression ratio is 64, the NMSE are respectively -12.76 dB and -11.21 dB, while the performance gain is still approximately 1.5 dB in terms of the NMSE. 

Moreover, considering that the CDL-A, CDL-B, and CDL-C models are all non-line-of-sight (NLoS) scenarios, we also conduct experiments using the CDL-D model, which simulates the line-of-sight (LoS) transmission channel \cite{3gpp38901}, to evaluate the performance of DiReNet in the LoS scenarios. Under the CDL-D model with a delay spread of 30 ns and a UE speed of 3 km/h, the NMSEs of DiReNet are -24.19 dB and -11.63 dB when the compression ratio $\sigma$ are respectively 8 and 64. The experimental results indicate that the proposed network exhibits satisfactory performance in the LoS scenarios.

\begin{table*}[t]
    \center
    \caption{Network Complexity Comparison}
    \setlength{\tabcolsep}{7.5mm} 
    \renewcommand{\arraystretch}{1} 
    \begin{tabular}{c|cccc|c}
    \Xhline{1.1pt}
    \multirow{2}*{Methods} &
    \multicolumn{4}{c|}{Compression ratio $\sigma$} & \multirow{2}*{Total param.}\\
    \cline{2-5} & 8 & 16 & 32 & 64 \\
    \hline CsiNet $\cite{CSI0}$ & 1.054M & 530K & 268K & 137K & 1.989M\\
    \hline CRNet $\cite{CRNet}$ & 1.054M & 530K & 267K & 136K & 1.987M\\
    \hline ACRNet-1x $\cite{ACRnet}$ & 1.054M & 529K & 267K & $\textbf{136K}$ & 1.986M \\
    \hline ACRNet-5x $\cite{ACRnet}$ & 1.062M & 538K & 276K & 145K & 2.021M\\
    \hline $\textbf{DiReNet}$ & $\textbf{695K}$ & $\textbf{390K}$ & $\textbf{236K}$ & 160K & $\textbf{1.481M}$\\
    \Xhline{1.1pt}
    \end{tabular}
    \label{net_complex}\vspace{-0.1cm}
\end{table*}

In our experiments, network complexity is measured by the number of parameters. As demonstrated in Table \ref{net_complex}, the proposed DiReNet has fewer parameters compared to other networks when the compression ratios are 8, 16, and 32. This can be attributed to the employment of several smaller FC layers in the proposed DiReNet, as opposed to a large FC layer employed in other networks. Furthermore, by reducing its reliance on FC layers and emphasizing the design of convolutional attention layers, DiReNet effectively enhances its CSI feature extraction capabilities while simultaneously decreasing the number of parameters.

\subsubsection{Network Complexity}

On the other hand, when the compression ratio is 64, the parameters in DiReNet is slightly larger than in other networks, while the former exhibits a significantly better NMSE performance than the latter. This is due to the fact that as the compression ratio increases, the number of parameters in FC layers decreases while the convolutional attention layers remain unchanged. In practical applications, different networks with various compression ratios need to be deployed, rendering the parameters with high compression ratios negligible compared to the total number of parameters. As shown in Table \ref{net_complex}, the total number of trainable parameters in DiReNet is significantly lower than in other advanced networks, representing a reduction of approximately one third.

\begin{figure}[t] 
\centering
    \includegraphics[width=0.65\linewidth]{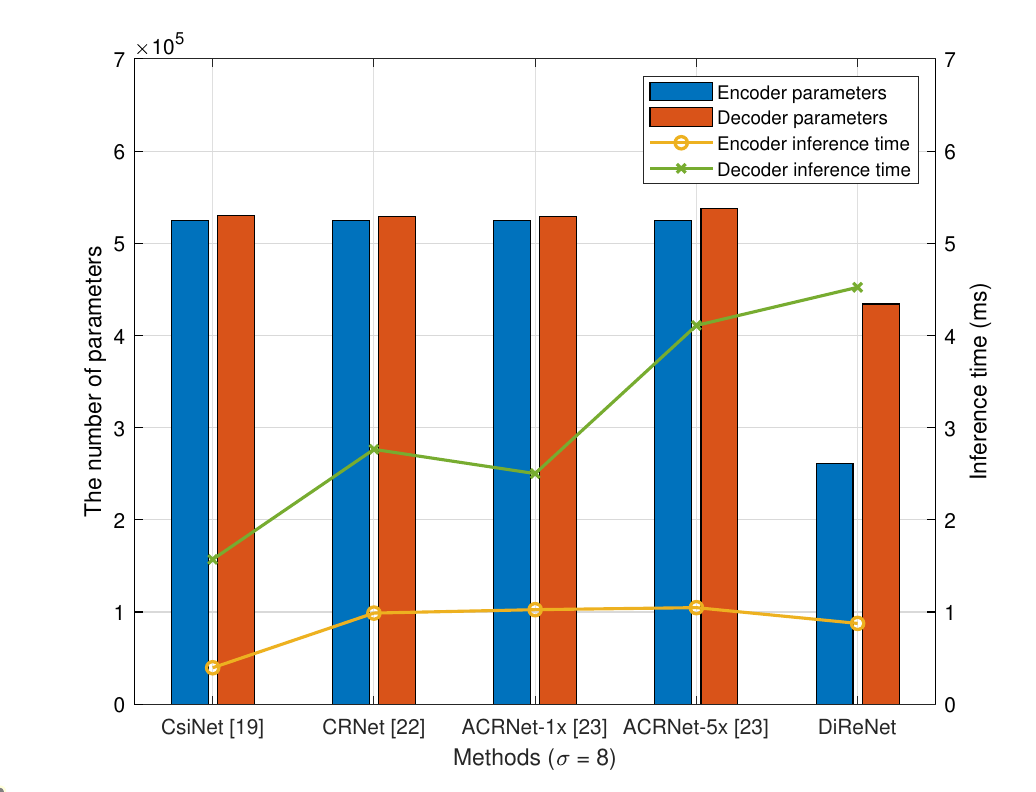}\vspace{-0.5cm}
    \caption{The parameters and inference time of encoder and decoder ($\sigma$ = 8).}
    \label{para_plot}\vspace{-0.1cm}
\end{figure}

In practical applications, the deployment of neural networks with various compression ratios requires a certain degree of memory and computational resources. Indeed, the majority of the parameters in our proposed DiReNet design are constituted by the decoder. As a result, the primary computation is offloaded to the high-performance BS. When the compression ratio is 8, we have provided a comparison of the parameters and inference time of both the encoder and decoder with other methods, as shown in Fig. \ref{para_plot}. It is observed that the proposed DiReNet reduces the number of parameters in both the encoder and decoder, thereby making the memory consumption on UE wireless devices acceptable. Besides, the inference time is also reduced in the encoder at the wireless device, while the computational cost is more effectively offloaded to the BS.

\subsection{Performance with Quantization Feedback}
In the proposed quantization scheme, distinct quantizers and quantization-levels are utilized for different vectors. By adopting uniform quantization, the quantization-level of the polarization-shared quantizer is set to $Q_\text{SA}$ and the polarization-specific quantization-level is $Q_\text{SP}$. The total feedback bits $B$ of a CSI matrix with dimensions $N_\text{s} \times N_\text{t}$ can be expressed as
\begin{equation}
B = \frac{2N_\text{s}N_\text{t}}{3\sigma} \left( Q_\text{SA} + 2 Q_\text{SP} \right).
\label{feed_bt}
\end{equation}

Table \ref{quan_bits3} presents the total quantization feedback bits under various compression ratios and quantization-level configurations. It can be observed that the total number of bits remains consistent when employing different quantization-level configurations for polarization-shared and polarization-specific vectors. Specifically, when the configurations of $Q_\text{SA}$ and $Q_\text{SP}$ are set to (4, 4), (2, 5), and (6, 3), respectively, the total number of feedback bits remains constant. Furthermore, the total number of feedback bits for the quantization configurations of (3, 3), (1, 4), and (5, 2) are equivalent.

\begin{table}[t]
    \center
    \caption{Comparison of Feedback Bits}
    \setlength{\tabcolsep}{4mm}
    \renewcommand{\arraystretch}{1}
    \begin{tabular}{c|c|c}
    \Xhline{1.1pt} 
    \multirow{2}{*}{$\sigma$} & \multicolumn{2}{c}{($Q_\text{SA}$, $Q_\text{SP}$)} \\ 
    \cline{2-3} & (3, 3), (1, 4), (5, 2)&  (4, 4), (2, 5), (6, 3) \\ 
    \hline  8  & 768 & 1024\\
    \hline  16 & 384 & 512 \\
    \hline  32 & 192 & 256\\
    \hline  64 & 96 & 128 \\
    \Xhline{1.1pt}
    \end{tabular}
    \label{quan_bits3}
    \vspace{-0.1cm}
\end{table}

\begin{figure}[t]
	\centering
	\includegraphics[width=0.65\linewidth]{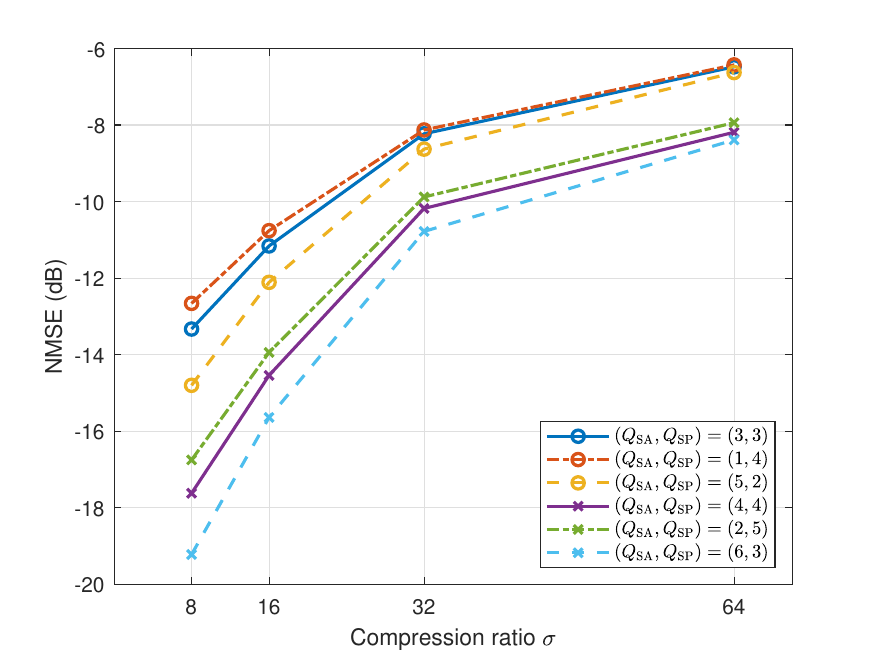}\vspace{-0.2cm}
	\caption{NMSE performance under different quantization configurations.}
	\label{quan_nmse}
    \vspace{-0.1cm}
\end{figure}

Fig. \ref{quan_nmse} illustrates the NMSE performance for various quantization-level configurations. In particular, an increase in the quantization-level of the shared vector results in an improvement in NMSE performance, while a decrease in the quantization-level of the shared vector leads to a slight decline in NMSE performance. This indicates that for dual-polarized CSI with high correlations, it is beneficial to allocate a higher quantization-level to shared information relative to specific information due to the reuse of the shared vector. Moreover, for dual-polarized CSI with varying degrees of correlation, the quantization-levels of both the shared and specific vectors can be flexibly configured to achieve enhanced performance.

\begin{figure*}[t]  
    \begin{center} 
        \subfigcapskip=-5pt
        \subfigure[$\sigma$ = 8.]{\includegraphics[width=0.49\linewidth]{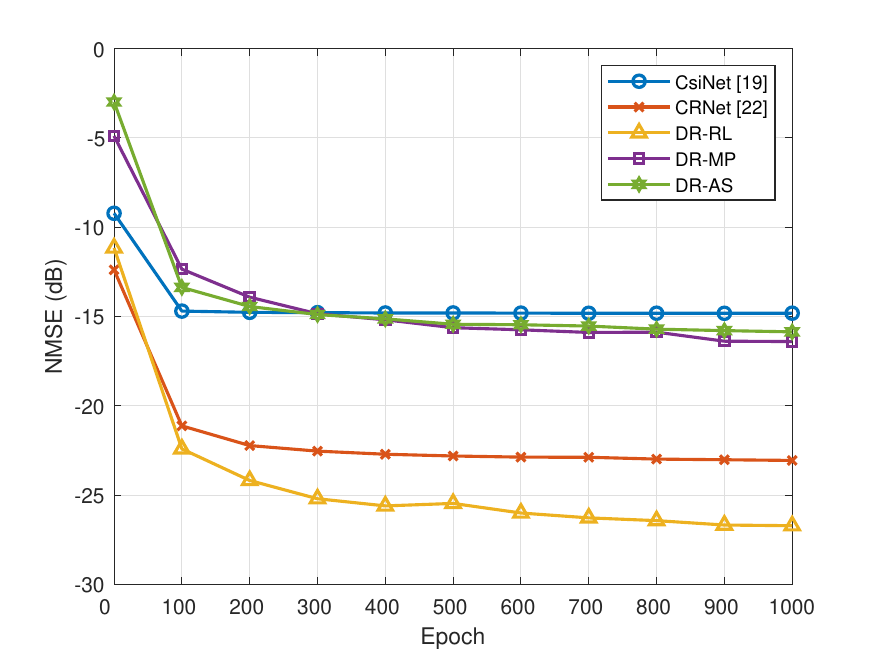}}\hspace{-5pt}
        \subfigure[$\sigma$ = 16.]{\includegraphics[width=0.49\linewidth]{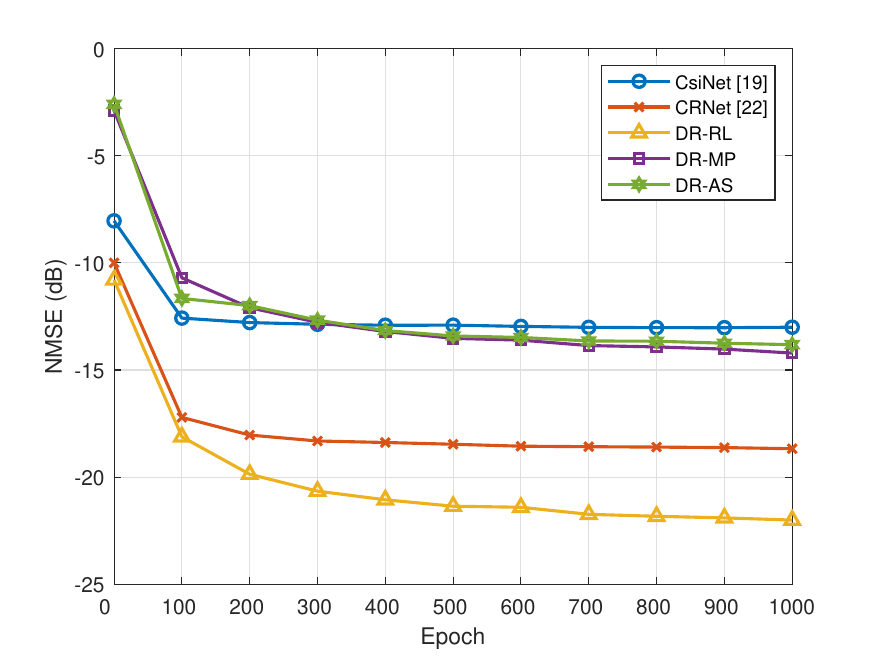}}\vspace{-5pt}
        \subfigure[$\sigma$ = 32.]{\includegraphics[width=0.49\linewidth]{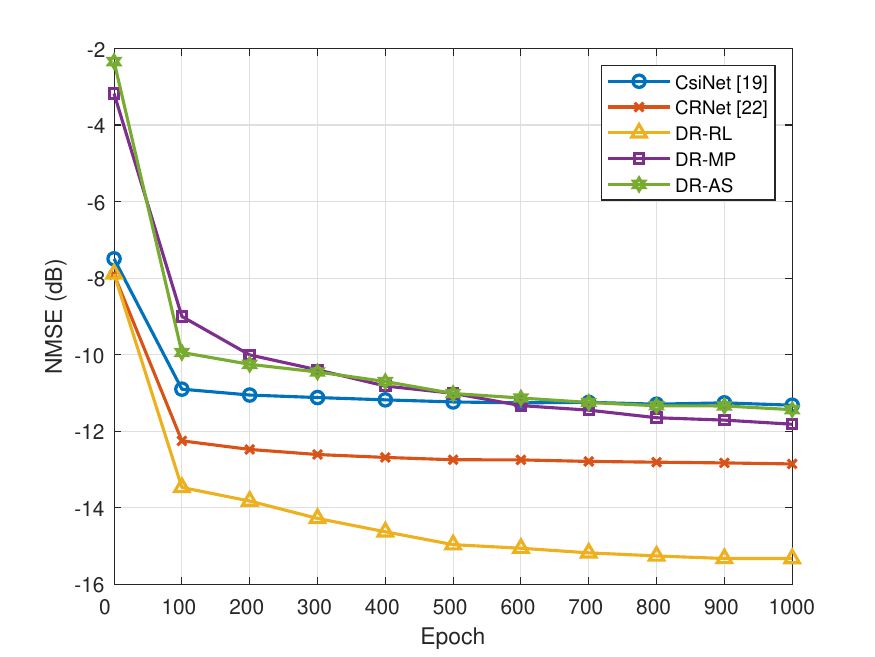}}\hspace{-5pt}
        \subfigure[$\sigma$ = 64.]{\includegraphics[width=0.49\linewidth]{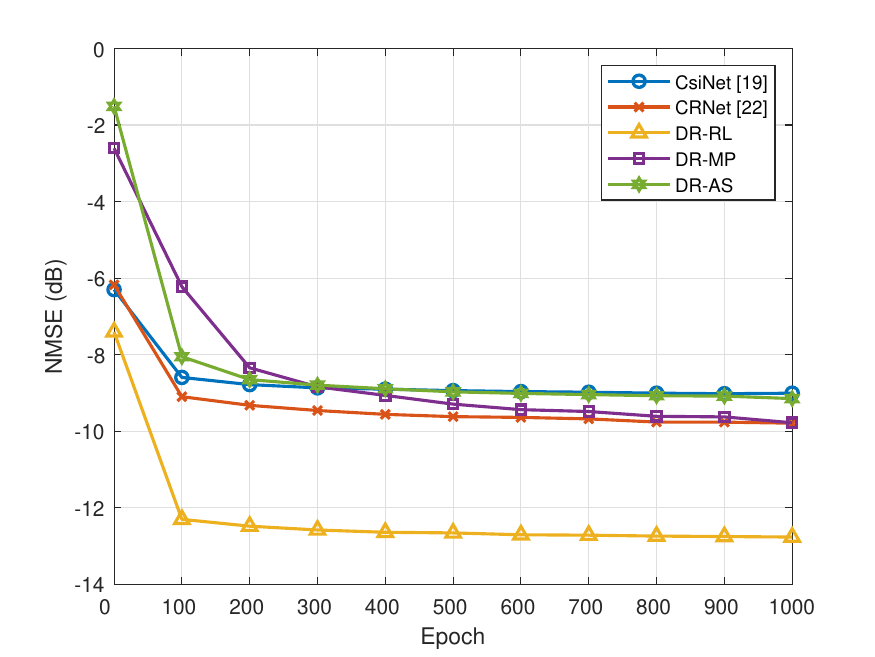}}\vspace{-0cm}
        \caption{NMSE performance of existing methods, the proposed method, and its two types of variants for ablation study.}\vspace{-0.3cm}
        \label{abo}
    \end{center}
\end{figure*}

\subsection{Impact of Disentangled Representation Learning}

In this subsection, we examine the effectiveness of the extracted disentangled representations by conducting ablation experiments. In one ablation, the average of the vertical and horizontal CSI magnitude is directly utilized as the shared representation, while the vertical and horizontal phases are employed as the vertical and horizontal specific representations, respectively. Additionally, in another ablation, the absolute value (ABS) of the channel coefficients is considered as the shared representation, while the signs are regarded as the specific representations. In this ablation, since the real and imaginary parts of CSI are input into the network twice, the final result is calculated from the average of the two results.

We compare existing DL-based networks with three proposed approaches, including the proposed DiReNet with disentangled representation learning (DR-RL), the ablation of magnitude and phase inputs (DR-MP), and the ablation of ABS and sign inputs (DR-AS). Fig. \ref{abo} shows that the proposed DiReNet outperforms its ablations and existing methods. This suggests that disentangled representation learning can effectively extract representations of shared and specific information. In the absence of disentangled representation learning, the performance of ablation experiments deteriorates. The representations obtained through disentangled representation learning exhibit enhanced performance compared to the direct utilization of certain channel parameters as representations.

\begin{table*}[t]
    \center
    \caption{NMSE(dB) Comparison of Different MI Distance}
    \setlength{\tabcolsep}{5mm}
    \renewcommand{\arraystretch}{1} 
    \begin{threeparttable}
    \begin{tabular}{c|ccccccc}
    \Xhline{1.1pt}
    \multirow{2}*{Compression ratio $\sigma$} &
    \multicolumn{7}{c}{MI distance (bit)*}\\
    \cline{2-8} & -100 & -50 & -25 & 0 & 25 & 50 & 100\\ 
    \hline 8 & -24.92 & -25.43 & -26.02 & $\textbf{-26.74}$ & -26.24 & -25.94 & -24.96\\ 
    \hline 16 & -19.52 & -20.82 & -21.37 & $\textbf{-21.97}$ & -21.83 & -21.51 & -20.60\\ 
    \hline 32 & -14.75 & -14.83 & -15.03 & $\textbf{-15.40}$ & -15.18 & -14.88 & -14.82\\ 
    \hline 64 & -11.31 & -11.99 & -12.59 & $\textbf{-12.76}$ & -12.63 & -12.57 & -11.97\\  
    \Xhline{1.1pt}
    \end{tabular}
    \label{table_reg}
    \begin{tablenotes}
        \footnotesize
        \item *MI distance denotes the complexity of the polarization-shared representation $\mathbf{W}$ (the larger the distance, the higher the complexity of the information contained in the representation).
      \end{tablenotes}
    \end{threeparttable}
    \vspace{-0.3cm}
\end{table*}

\subsection{Impact of Mutual Information Regularization}
In this subsection, we investigate the effectiveness of the proposed MI regularization by adjusting the distance between $\mathcal{I}\left(\mathbf{H}_\text{v}, \mathbf{H}_\text{h}; \mathbf{W}\right)$ and $\mathcal{I}\left(\mathbf{H}_\text{v}; \mathbf{H}_\text{h}\right)$. As presented in Table \ref{table_reg}, we verify that the proposed MI regularization can learn the most sufficient and effective representations. On the one hand, when the difference between $\mathcal{I}\left(\mathbf{H}_\text{v}, \mathbf{H}_\text{h}; \mathbf{W}\right)$ and $\mathcal{I}\left(\mathbf{H}_\text{v}; \mathbf{H}_\text{h}\right)$ is negative, i.e., the complexity of the shared representation is less than the correlation between dual-polarized CSI, the NMSE performance deteriorates, indicating that the network has not learned an adequate representation to capture all the highly correlated channel information. On the other hand, when the difference is positive, i.e., the complexity of the representation is greater than the polarization correlations, the performance is reduced, indicating that an effective representation has not been obtained. In summary, the proposed MI regularization effectively guides the learned representation $\mathbf{W}$ to capture all necessary correlated channel characteristics.

\subsection{Impact of Depth-wise and Width-wise Extensions}

In this subsection, we compare the NMSE performance with different depth-wise and width-wise extensions in the decoder. We also analyze the impacts of different depths and widths on the network, which reveals the tradeoff between NMSE performance and network complexity. 

\begin{figure*}[t]  
    \begin{center}  
        \subfigure[Depth-wise extensions.]{\includegraphics[width=0.49\linewidth]{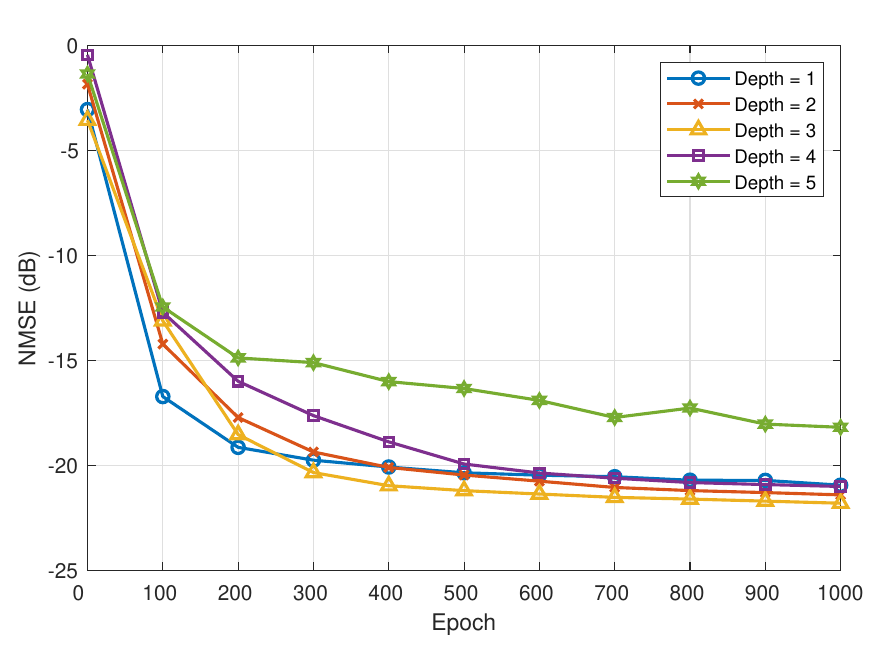}} 
        \subfigure[Width-wise extensions.]{\includegraphics[width=0.49\linewidth]{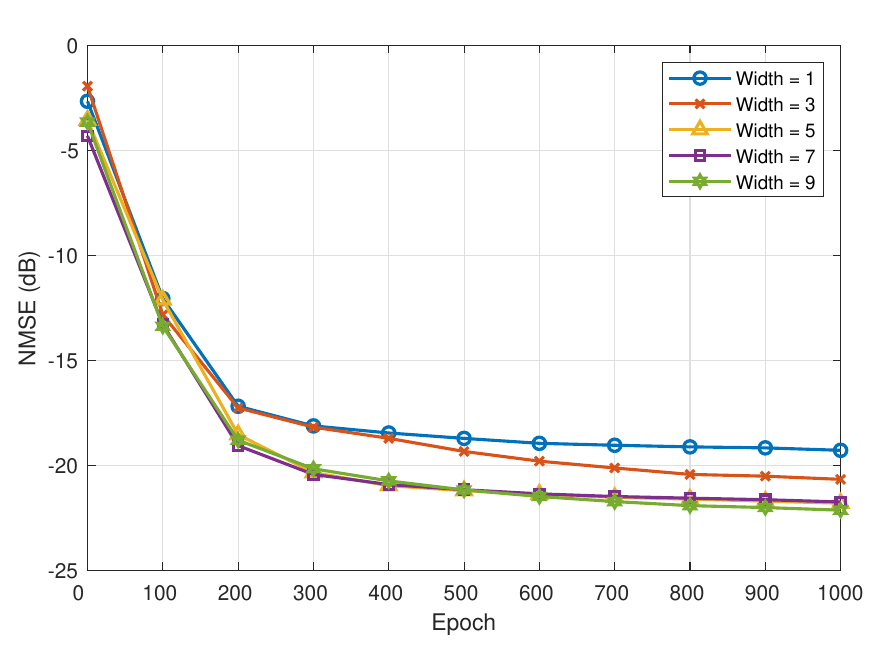}}\vspace{-0.1cm}
        \caption{Comparison of NMSE performance under different depth and width-wise extensions ($\sigma$ = 16).}\vspace{-0.4cm}
        \label{depth_width_ex}
    \end{center}
\end{figure*}

For the depth-wise extensions experiment, we fix the compression ratio at 16 and the width-wise extension at 5, while varying the depth-wise extensions as 1, 2, 3, 4, and 5, respectively. The results are presented in Fig. \ref{depth_width_ex}(a), we observe that although the networks with fewer extensions converge faster during the initial 200 training epochs, the final convergence performance is unsatisfactory. This is due to the limited learning capacity of shallower networks, despite their relative ease of training, to capture intricate details for accurate CSI recovery. Furthermore, we note a significant decrease in NMSE performance when the depth is excessively increased, potentially because an overly deep network leads to the loss of CSI information. Specifically, optimal performance is achieved with a depth of 3. This illustrates that DiReNet not only extracts the CSI feature effectively, but also avoids the performance degradation.

For the width-wise extensions experiment, we set the compression ratio to 16 and the depth-wise extension to 3, while the width-wise extensions are 1, 3, 5, 7, and 9, respectively, Fig. \ref{depth_width_ex}(b) shows the results. It is noteworthy that as the width of DiReNet is extended, it exhibits an enhanced capacity to extract a more diverse range of CSI features, culminating in a commensurate improvement in its NMSE performance. However, with an increase in the width, the training procedure becomes more challenging and the required number of network parameters increases linearly. Consequently, a width-wise extension of 5 is selected to strike a balance between NMSE performance and network complexity.

\subsection{Generalization Performance of the Proposed DiReNet}
\begin{figure}[t] 
\centering
    \includegraphics[width=0.65\linewidth]{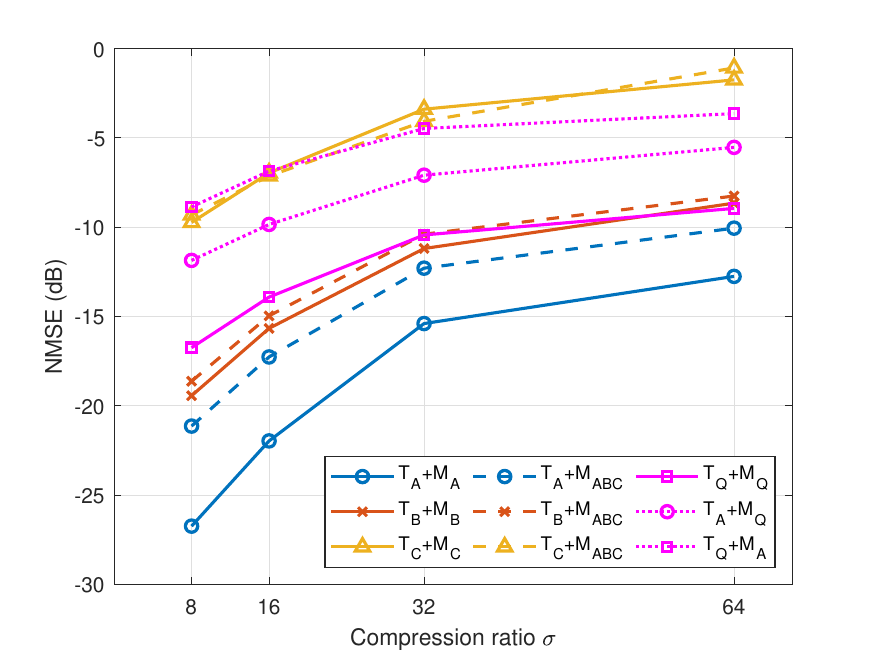}\vspace{-0.1cm}
    \caption{Generalization performance comparison. ($\text{T}_\text{A/B/C}$ denotes the testing channel data are CDL-A/B/C, respectively. $\text{T}_\text{Q}$ denotes the testing channel data are generated by QuaDRiGa. $\text{M}_\text{A/B/C}$ denotes the models are trained by the training channel data CDL-A/B/C, respectively. $\text{M}_\text{ABC}$ denotes the model is trained by three scenarios simultaneously. $\text{M}_\text{Q}$ denotes the models are trained by the QuaDRiGa training data.)}
    \label{general}
    \vspace{-0.3cm}
\end{figure}

In this subsection, we analyze the generalization performance of the proposed DiReNet in various channel scenarios. First, we train the proposed DiReNet on a single channel scenario and test its NMSE performance. As it can be observed from Fig. \ref{general}, the NMSE performance of CSI recovery in the CDL-A channel scenario is superior to that in the CDL-B and CDL-C channel scenarios. This suggests that increased user mobility and a greater delay spread contribute to more rapid variations in the channel, thereby increasing the complexity of the channel environment. As a result, the neural network faces increased challenges in accurately extracting features, compression, and recovery of the CSI.

Moreover, to demonstrate the generalization abilities of the proposed DiReNet, we train it utilizing data from all CDL-A, CDL-B, and CDL-C channels. We then compare the recovery performance across different channel scenarios with other instances of the DiReNet trained on a single channel scenario. When training data from all CDL-A, CDL-B, and CDL-C channels, to ensure fairness in the training process, we mix all 300,000 training data together and randomly select 100,000 samples to form the new training set. From Fig. \ref{general}, it is illustrated that the recovery accuracy of the testing CSI in the CDL-A scenario experiences a relatively significant decrease, while the accuracy reduction in the CDL-B and CDL-C scenarios for CSI recovery is marginal. Although the recovery accuracy of the neural network trained on different channel scenarios is not as high as on a single channel scenario, this method has demonstrated that our proposed DiReNet possesses generalization abilities and can adapt effectively to various channel scenarios.

\begin{figure}[t] 
\centering
    \includegraphics[width=0.65\linewidth]{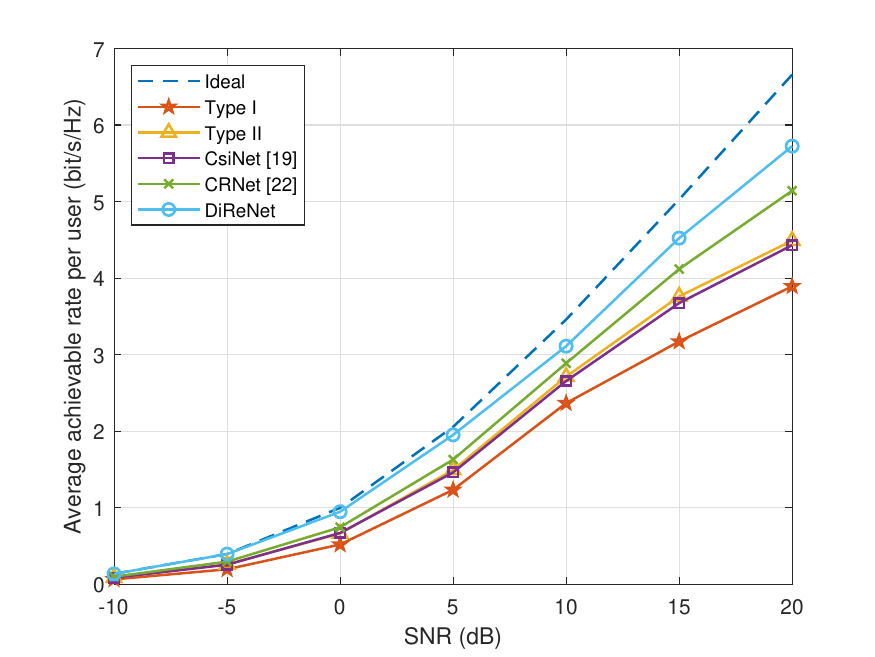}\vspace{-0.1cm}
    \caption{Comparison of average achievable rate per user performance under different methods.}
    \label{rate_add}
    \vspace{-0.2cm}
\end{figure}

To further evaluate the generalization of the network under various channels, additional experiments have been added to test unknown channel data with the trained network. We generate new simulation dataset using the QuaDRiGa model \cite{quadriga}. As depicted in Fig. \ref{general}, the experiments include testing the QuaDRiGa channel on a network trained by the CDL-A channel and testing the CDL-A channel on a network trained by the QuaDRiGa channel. It is observed that satisfactory performance is maintained even although there is some degradation in terms of the NMSE, when the trained network is applied to unknown channels.

\subsection{Achievable Rate Performance of the Proposed DiReNet}

In this subsection, we conduct experiments on the achievable rate performance of different DL-based and codebook-based CSI feedback methods to evaluate the end-to-end precoding performance. In our experiment, we exploit the recovered CSI matrix to obtain the zero forcing (ZF) precoding matrix for 4 users and assume that each user has the same distance from the transmitter. Then, we compare the achievable rate per user performance of three DL-based methods, including CsiNet \cite{CSI0}, CRNet \cite{CRNet} and our proposed DiReNet, along with two codebook-based methods, i.e., Type I and Type II \cite{3gpp_2}. As observed from Fig. \ref{rate_add}, our proposed DiReNet can achieve the highest rate and outperforms the other methods across the entire considered signal-to-noise ratios (SNRs) at the receiver. This can be attributed to the fact that our proposed DiReNet can reduce information redundancy and extract features more effectively.

\section{Conclusion}
In this paper, we proposed a deep CSI feedback network named DiReNet based on disentangled representation learning. Unlike existing DL-based methods, which treat the CSI matrix as an image directly, our approach exploited the disparity between the characteristics of natural images and CSI for the design of a parametric encoder and decoder. Particularly, for dual-polarized CSI, the proposed DiReNet utilized the inherent correlation between different polarization directions to enhance network performance. In the encoder, the correlated dual-polarized CSI was decomposed into the polarization-shared and polarization-specific information. In the decoder, the vertical polarization CSI was recovered by combining the shared information with the vertical specific information, while the horizontal polarization CSI was recovered by incorporating the shared information with the horizontal specific information. Additionally, our approach allowed for the implementation of different quantization schemes and the design of decoder extensions. Experimental results showed that the proposed DiReNet is superior to existing methods in terms of both NMSE performance and network complexity.

Although the DL-based methods have shown promising results, some challenging tasks remain to be solved in future work. In practice, deploying multiple models for various channel scenarios inevitably leads to an increase in network complexity. Therefore, how to further reduce the complexity of networks by more advanced DL-based models is still a topic worth studying. On the other hand, how to devise a network with sufficient generalization ability, which can leverage the inherent characteristics of CSI across all scenarios is an interesting and challenging research topic.

\bibliographystyle{IEEEtran}
\bibliography{ref.bib}

\end{document}